\documentclass[format=acmsmall, review=false, screen=true]{acmart}

\usepackage{booktabs} % For formal tables
\usepackage{todonotes}
\usepackage{color,soul}
\usepackage{array}
\usepackage{subfigure}
\usepackage{multirow}

\usepackage[ruled]{algorithm2e} % For algorithms

\SetAlFnt{\small}
\SetAlCapFnt{\small}
\SetAlCapNameFnt{\small}
\SetAlCapHSkip{0pt}
\IncMargin{-\parindent}

\acmJournal{TOIS}
\acmVolume{9}
\acmNumber{4}
\acmArticle{39}
\acmYear{2018}
\acmMonth{1}
\copyrightyear{2018}
\setcopyright{acmlicensed}
\acmDOI{0000001.0000001}

\received{July 2017}
\received[revised]{November 2017}
\received[accepted]{January 2018}

\begin{document}

\setstcolor{red}

% Title portion. Note the short title for running heads 
\title[General-Purpose Embeddings for User and Location Modelling]{Unsupervised Learning of Parsimonious General-Purpose Embeddings for User and Location Modelling}  
\author{Jing Yang}
\orcid{0000-0002-2565-0574}
\affiliation{%
  \institution{ETH Zurich}
  \streetaddress{Raemistrasse 101}
  \city{Zurich}
  \state{}
  \postcode{8092}
  \country{Switzerland}}
\email{jing.yang@inf.ethz.ch}
\author{Carsten Eickhoff}
\affiliation{%
  \institution{ETH Zurich}
  \streetaddress{Raemistrasse 101}
  \city{Zurich}
  \state{}
  \postcode{8092}
  \country{Switzerland}}
% \email{carsten.eickhoff@inf.ethz.ch}
% \author{Ting Yan}
% \affiliation{%
%   \institution{Eaton Innovation Center}
%   \city{Prague}
%   \country{Czech Republic}}
% \author{Tian He}
% \affiliation{%
%   \institution{University of Minnesota}
% \author{Chengdu Huang}
% \author{John A. Stankovic}
% \author{Tarek F. Abdelzaher}
% \affiliation{%
%   \institution{University of Virginia}
%   \department{School of Engineering}
%   \city{Charlottesville}
%   \state{VA}
%   \postcode{22903}

\begin{abstract}
Many social network applications depend on robust representations of spatio-temporal data. In this work, we present an embedding model based on feed-forward neural networks which transforms social media check-ins into dense feature vectors encoding geographic, temporal, and functional aspects for modelling places, neighborhoods, and users. We employ the embedding model in a variety of applications including \textit{location recommendation}, \textit{urban functional zone study}, and \textit{crime prediction}. For \textit{location recommendation}, we propose a \textbf{S}patio-\textbf{T}emporal \textbf{E}mbedding \textbf{S}imilarity algorithm (STES) based on the embedding model. 

In a range of experiments on real life data collected from Foursquare, we demonstrate our model's effectiveness at characterizing places and people and its applicability
% efficacy 
in aforementioned problem domains. 
% In particular, we significantly increase recommendation and prediction performance as compared to the state of the art.
Finally, we select eight major cities around the globe and verify the robustness and generality of our model by porting pre-trained models from one city to another, thereby alleviating the need for costly local training. 
\end{abstract}




\setcopyright{acmcopyright}
\acmJournal{TOIS}
\acmYear{2018} \acmVolume{1} \acmNumber{1} \acmArticle{1} \acmMonth{1} \acmPrice{\$15.00}


% \thanks{
% %We thank Zhiyuan Cheng for providing us with the check-in data set, and we thank Bo Hu, Min Xie, and Hongzhi Yin for sharing their location recommendation algorithm details. We also thank Jan D\"orrie, Zack Zhu, Hangxin Lu, and Hongyu Xiao for their insightful discussions. 

%   Author's addresses: J. Yang {and} C. Eickhoff, Department of Computer Science, ETH Z\"urich, R\"amistrasse 101, 8092 Z\"urich, Switzerland}

\maketitle

\renewcommand{\shortauthors}{J. Yang et al.}

\section{Introduction} \label{sec:intro}

Spatial social network applications and services, such as transport network design, location-based services, urban structure learning, and crime prediction, normally involve two important issues: understanding residents' real-time activities and accurately describing urban spaces~\cite{scellato2011exploiting, gerber2014predicting, church2000transport}. Towards the former, researchers usually rely on check-in data from social network platforms (\textit{e.g.}, Twitter, Foursquare, and Facebook) or specifically designed surveys, which cover a significant number of users in the form of check-ins and comments about points-of-interest (POIs). For the latter aspect, place annotation approaches are required. 

Among approaches to annotate places, a common method is to simply employ categorical labels such as \textit{Home}, \textit{Restaurant}, and \textit{Shop}~\cite{he2016spatial, krumm2013placer, sarda2016semantic, ye2011semantic}. Although straightforward, it is unclear whether such discrete tags offer sufficient flexibility and descriptive power for modelling the complexity of urban landscapes.  

% In this work, we represent locations by means of embedding vectors in a semantic space. Our framework solely relies on social media check-ins and the embedding model is built conceptually based on the \textit{Word2Vec} technique~\cite{mikolov2013distributed}. Aiming to annotate locations in terms of temporal, geographic, and functional aspects, we decompose the time, the locations, and the venue functions of check-in records into virtual ``\textit{words}''. We then treat check-in sequences as ``\textit{sentences}'' and regard activity profiles of a neighborhood or a user as ``\textit{documents}''.
In this work, we instead represent places by means of embedding vectors in a semantic space. Aiming to annotate places in terms of temporal, geographic, and functional aspects, we extract the time, location, and venue function from check-in records and train our model in the context of check-in sequences which originate from an individual user or neighborhood.
% we train our embedding model based on the time, the locations, and the venue functions of social media check-in records in the contexts of check-in sequences originating from a user or from within a neighborhood. 

In comparison with the traditional discrete method, our approach describes places in a continuous manner and preserves more information about people's real behavior as well as places' day-to-day usage patterns. For instance, in the case of label annotation, three food related places which serve Chinese breakfast, pizza, and sushi, respectively, may all be labeled as \textit{Restaurant} but their features in food type, active hours, and location may vary dramatically. In the course of this article, we will show how our embedding model represents such within-class variance in a natural way. As embedding vectors are learnt from people's real-time check-ins, we also leverage them in user representation to reflect people's activity patterns and interests.

The embedding model is an accurate descriptor of places and users in terms of geographic and functional affinity, activity preferences, and daily schedules. As a consequence, it can be applied in a wide range of settings. In this paper, we consider three practical applications: \textit{location recommendation}, \textit{urban functional zone study}, and \textit{crime prediction}. Our empirical investigation is driven by five research questions:

\begin{itemize}
    \item RQ1. How well does the embedding model differentiate locations and users along temporal, geographic, and functional aspects?
    \item RQ2. How does our location recommendation algorithm STES compare to state-of-the-art methods?
    \item RQ3. How to define and visualize urban functional zones using the proposed model?
    \item RQ4. How well can the model predict typical urban characteristics?
    \item RQ5. With what generalization error can an embedding model trained in one city be transferred to other cities?
\end{itemize}

By answering these research questions, we make three novel contributions:

\begin{itemize}
\item We present an \textbf{unsupervised} spatio-temporal embedding scheme based on social media check-ins. Trained with monthly check-in sequences, the model shows wide applicability for tasks ranging from social science problems to personalized recommendations. 
\item Based on this model, we propose the \textbf{STES} algorithm that recommends locations to users. Compared with state-of-the-art recommendation frameworks, we can achieve an improvement up to $30\%$.
\item The model shows strong \textbf{robustness} and \textbf{generality}. Once trained in a representative city, it can be directly utilized in other cities with only slight generalization errors ($<3\%$).   
\end{itemize}   

The remainder of this article is structured as follows: Section~\ref{sec:related} summarizes existing literature on check-in embedding learning as well as state-of-the-art works in \textit{location recommendation}, \textit{urban functional zone study}, and \textit{crime prediction}. 
% gives an overview of related work on representation learning of spatio-temporal data for location recommendation and crime prediction. 
Section~\ref{sec:methodology} describes our embedding model and the STES location recommendation algorithm. Section~\ref{sec:experiments} empirically evaluates the performance of our model on a number of tasks, comparing to a range of competitive performance baselines. Finally, Section~\ref{sec:conclusion} concludes with a summary of our findings and a discussion of future work.

\section{Related Work}\label{sec:related}

% In this section, we first review \textit{Word2Vec} and its applications on basis of social networks. Then, we introduce the state-of-the-art in our two application domains, location recommendation and crime prediction.

In this section, we first discuss the progress of place annotation using check-in data. We then review embedding learning techniques and their applications on the basis of social networks. Finally, we introduce the state of the art in our three application domains.

\subsection{Place Annotation Using Check-in Data}
Among place annotation works that involve social media or survey data, most studies utilize existing venue category tags and formulate the problem as a prediction or clustering task. Sarda \textit{et al.}~\cite{sarda2016semantic} propose a spatial kernel density estimation based model on the basis of the 2012 Nokia Mobile Data Challenge (MDC) \cite{laurila2012mobile} to label an unknown place with one of 10 semantic tags (\textit{e.g.,} \textit{Home}, \textit{Transport Related}, \textit{Shop}). He \textit{et al.}~\cite{he2016spatial} design a topic model framework which takes user check-in records as input and annotates POIs with category-related tags from Twitter and Foursquare. Noulas \textit{et al.}~\cite{noulas2011exploiting} represent areas by counts of inner check-in venue categories and further implement clustering. Krumm and Rouhana \cite{krumm2013placer} propose an algorithm to classify locations into 12 labels (\textit{e.g.}, \textit{School}, \textit{Work}, \textit{Recreation Spots}) based on visitor demographics, time of visit, and nearby businesses of the locations. Ye \textit{et al.}~\cite{ye2011semantic} propose a support vector machine based algorithm to annotate places with 21 category tags. As mentioned earlier, such discrete usage of category labels carries insufficient descriptive power for modelling users and the real activities in urban spaces. Therefore, a continuous vector representation in semantic space is proposed in our work.

\subsection{Embedding Learning Techniques and Applications}
% Intended for natural language processing, \textit{Word2Vec}~\cite{mikolov2013distributed} processes textual documents in an unsupervised manner and learns embedding vectors for words and phrases based on their context of occurrence in text. Due to its efficiency in capturing the contextual correlation of items, researchers have been applying the technique for E-commerce product recommendation~\cite{phidistributed2016diudiu}, network embedding~\cite{perozzi2014deepwalk}, user profiling~\cite{tang2015learning, tang2015user}, social media prediction tasks~\cite{volkovaaccount2016lala, wijeratne2016word}, and many others. However, when employed in social network context, the model is still mainly used on texts such as tweets and Yelp reviews. Our analogy is to treat check-in sequences as virtual ``\textit{sentences}'' and users or neighborhoods as ``\textit{documents}''. Consequently, correlations of contextual locations and activities can be better modelled.
In recent years, due to promising performance in capturing the contextual correlation of items, approaches to embedding items in Euclidean spaces have become popular and have been applied in a variety of domains, including E-commerce product recommendation~\cite{phidistributed2016diudiu}, network classification~\cite{perozzi2014deepwalk}, user profiling~\cite{tang2015learning, tang2015user}, etc. Social media contexts are active fields of application. Wijeratne \textit{et al.}~\cite{wijeratne2016word} embed words in tweet texts and twitter profile descriptions into vectors for gang member identification. Tang \textit{et al.}~\cite{tang2014learning} learn embeddings of sentiment-specific words in tweets for sentiment classification. Lin \textit{et al.}~\cite{lin2017structured} develop a matrix sentence embedding framework and adopt it in Yelp reviews for user sentiment analysis. In these cases, however, embedding techniques are straightforward applications to textual documents. In our work, we develop an analogous situation by treating check-in sequences as virtual sentences. Consequently, correlations of contextual locations and activities can be better modelled.  

\subsection{Location Recommendation}
As the most common performance benchmark for spatial models~\cite{cheng2012fused, ye2010location}, location recommendation has been popular in recent years in studies on location-based social networks (LBSNs). The most basic location recommendation approaches are content based, relying solely on properties of users and locations~\cite{massa2007trust, tavakolifard2012social}. The idea is to explore user and location features, then make recommendations based on their similarities and past preferences. Another popular branch of approaches employs matrix factorization (MF)~\cite{koren2009matrix, salakhutdinov2011probabilistic} and its derivative methods~\cite{cheng2012fused, cheng2013you, lian2014geomf, liu2013learning}. MF-related methods aim to represent users and items in matrices and recommend locations based on row-to-row correlations. Methods based on topic models (TM)~\cite{blei2003latent} and Markov models (MM)~\cite{mathew2012predicting} also perform well in recommendation tasks. Here, geographic and temporal information are often included as additional evidence~\cite{hong2012discovering, hu2013spatio, kurashima2013geo, gao2013exploring, li2015rank, zhang2014lore}. There exist several topic-model based location recommendation methods which consider geographic, temporal, and even venue-specific semantic information~\cite{wang2015geo, yin2016adapting, yin2016joint, yin2014lcars, yin2015joint, wang2016spore}. However, their problem setting is different. Instead of predicting the next place which a user might visit, they focus on recommending a nearby venue where a user has never been before. To optimize such a recommendation, they further introduce the concepts of \textit{home users} and \textit{visitor users} to accommodate for various visitation constellations.

Recently, some recommendation methods~\cite{ozsoy2016word, liuexploring2016dada, zhao2016gt} involve neural network embedding techniques in their processing schemes. However, they only focus on geographic location modelling but ignore other information (\textit{e.g.}, venue function, check-in time) associated with each check-in. Consequently, these models hold very limited applicability beyond the immediate context of location recommendation that they were designed for. There also exist some recommendation approaches involving LSTM recurrent neural networks~\cite{song2016multi, li2016tweet}. However, such methods fall into the class of supervised learning, requiring costly annotations that are not always available. A recent related work proposed a spatial-aware hierarchical collaborative deep learning model for location recommendation~\cite{yin2017spatial} which demonstrates promising performance in handling cold start issue. However, like the works of topic modelling approaches above, the problem scenario is to recommend POIs which have never been visited before. Additionally, these methods exploit the semantic representation of POIs in a supervised and task-guided way. In contrast, we develop an unsupervised learning approach including various types of check-in information which can be utilized in a wide range of settings by giving a robust representation of each place without requiring supervised label information. Section~\ref{sec:experiments} will highlight the resulting performance differences both in the context of recommendation as well as other tasks. 

\subsection{Urban Functional Zone Study}
Usually, urban functional zone partitioning is purely based on geographic data from GIS and government data sets \cite{tian2010spatial, yajun2006functionality}. In recent years, researchers began to incorporate crowdsourced activity data from social media, public surveys, and traces of personal mobility into such studies. Xing \textit{et al.}~\cite{xing2014traffic} use mobile billing data to cluster urban traffic zones. Yuan \textit{et al.}~\cite{yuan2015discovering} design a topic model and use POIs and taxi trajectories together with public transit data to cluster neighborhoods in Beijing into six functional zones such as \textit{Education}, \textit{Residence}, and \textit{Entertainment}. Zhu \textit{et al.}~\cite{zhu2015learning} leverage 2014 Puget Sound travel survey data\footnote[1]{http://www.psrc.org/data/transportation/travel-surveys/2014-household}, Foursquare POIs, and Twitter temporal features to characterize and cluster neighborhoods into four functional areas \textit{Shopping}, \textit{Work}, \textit{Residence}, and \textit{Mixed functionalities}. Cranshaw \textit{et al.}~\cite{cranshaw2012livehoods} propose a spectral clustering method to directly group POIs based on their check-in information.
In our work, we simply utilize the check-in vectors trained by our embedding model to characterize neighborhoods and further cluster urban functional zones. We show that our method is capable of encoding people's daily activity patterns to discover the true underlying usage of urban spaces. 

\subsection{Crime Prediction}
Crime prediction is a common social science issue. Iqbal \textit{et al.}~\cite{iqbal2013experimental} rely on demographic features such as population, household income, and education level as input to predict regional crime rates on a three-point scale. Gerber~\cite{gerber2014predicting} and Wang \textit{et al.}~\cite{wang2012automatic} design topic models based on Twitter posts to predict criminal incidences. In this work, we demonstrate that people's day-to-day activity patterns are strong indicators of crime occurrence. Based on the proposed embedding model, we characterize neighborhoods with check-in vectors. Acting as features, these vectors perform well in future crime rate and crime occurrence prediction. 

\section{Methodology}\label{sec:methodology}

In this section, we begin by introducing the problem domain. Then, we describe the check-in embedding model. Finally, we propose the model-based location recommendation algorithm STES. 

\subsection{Scenario}
% As the only input data, social media check-ins are first grouped into ``\textit{words}'', ``\textit{sentences}'', and ``\textit{documents}'' for embedding training. Important definitions are proposed as follows. 
We first process social media check-ins in terms of their temporal, geographic, and functional aspects. Important definitions are proposed as follows.

\textbf{Check-ins.} A check-in is defined as a tuple $c = <u,f,t,l>$ which depicts that a user $u$ visits a location $l$ at time $t$, where $f$ demonstrates the functional role of the visited venue. To process the raw time data $t$, motivated by a natural reflection of daily routines and the data set density, we discretize $t$ in 10 functional timeslots, namely, \textit{Morning}, \textit{Noon}, \textit{Afternoon}, \textit{Evening}, \textit{Night}, \textit{WeekendMorning}, \textit{WeekendNoon}, \textit{WeekendAfternoon}, \textit{WeekendEvening}, and \textit{WeekendNight}. Discretization thresholds are listed in Table \ref{tab:correspondence}. 
% Timeslots have varying lengths to reflect daily routines. 
\textit{Morning} is the time bucket when people start a day and work before lunch. \textit{Noon} is the lunch break time. \textit{Afternoon} refers to working hours after lunch. \textit{Evening} is for dinner and after-work activities. \textit{Night} corresponds to the sleeping hours. As daily activities on weekends are usually different from those on weekdays (\textit{e.g.}, \textit{WeekendAfternoon} is often for leisure instead of work on weekday \textit{Afternoon}), we additionally define a set of timestamps on weekends that has the same temporal correspondence. To represent a check-in's functional role $f$, we utilize the popular Foursquare hierarchy of venue categories\footnote[2]{https://developer.foursquare.com/categorytree}. As for the location $l$, we leverage the unique id of each place. 
%A check-in's functional role $f$ is described following the popular Foursquare venue categories\footnote[2]{https://developer.foursquare.com/categorytree}.

\begin{table}
% \scriptsize
  \caption{Time Discretization}
  \label{tab:correspondence}
  \begin{tabular}{lc}
    \toprule
    Timeslot Tag&Time\\
    \midrule
    Morning/WeekendMorning & 6:00AM-10:59AM \\
    Noon/WeekendNoon & 11:00AM-1:59PM \\
    Afternoon/WeekendAfternoon & 2:00PM-5:59PM \\
    Evening/WeekendEvening & 6:00PM-9:59PM \\
    Night/WeekendNight & 10:00PM-5:59AM \\
  \bottomrule
\end{tabular}
\end{table}

For each check-in, only its function $f$, time $t$, and location $l$ are combined and trained to obtain the embedding vectors. There exist around 400 functional roles, 10 timeslots, and thousands of unique locations. Due to the large amount of locations, simply concatenating all these three aspects into one check-in word would result in a relatively sparse distribution of data points given the scale of available data. Therefore, to achieve a good balance, for each check-in record,
% Considering the check-in meaning, check-in size, and data set scale, for each check-in record, 
we concatenate functional role $f$ and timeslot $t$ as its \textit{feature word} (\textit{e.g.}, \textit{"Bar\_Evening"}), and use the unique id of location $l$ as its \textit{location word} (\textit{e.g.},\textit{"423e0e80f964a52044201fe3"}).

\begin{figure}
\includegraphics[width = 0.7\textwidth]{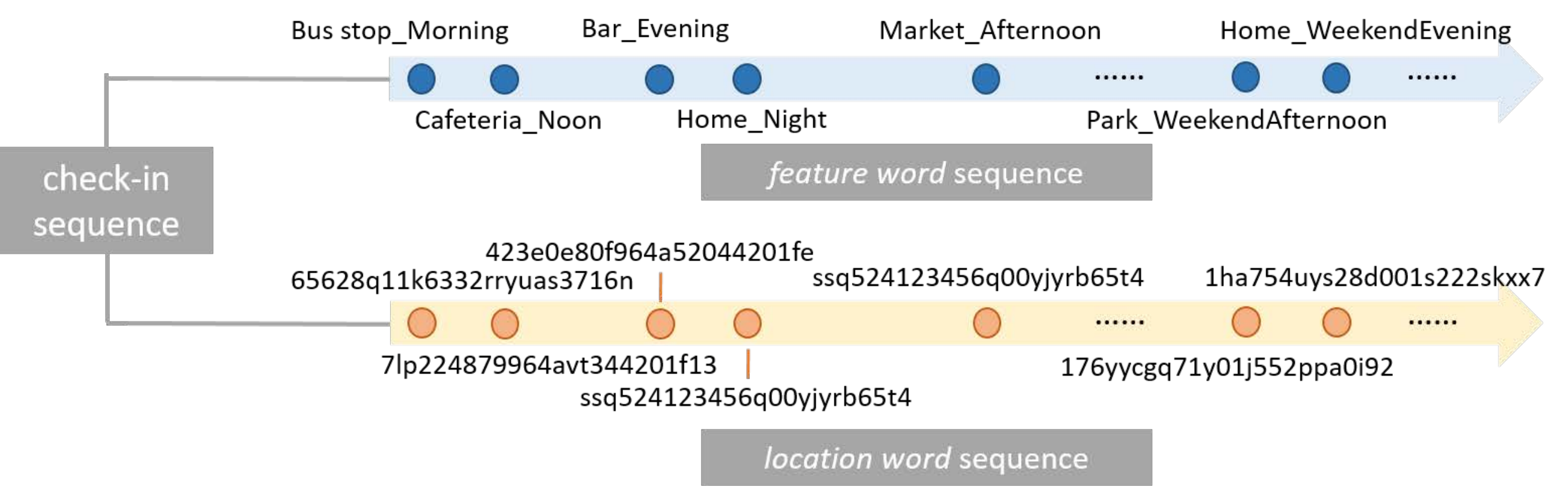}
\caption{A check-in sequence is composed of a \textit{feature word} sequence and a \textit{location word} sequence. Each dot represents a unique check-in. The dots are distributed unevenly to reflect the varying time spans between check-ins.}
\label{fig:sentence}
\end{figure}

\textbf{Check-In Sequences.} Check-in sequences are described by two parallel sub-sequences: a \textit{feature word} sequence and a \textit{location word} sequence. Words in these two sequences are one-to-one correspondent. Both the \textit{feature word} sequence and \textit{location word} sequence are chronological orderings of check-ins in one month of the profile of a user $u$ or a neighborhood $n$. Figure~\ref{fig:sentence} shows a check-in sequence example.
% shows a \textit{feature word} sequence example.

% % \stepcounter{figure}
% \begin{figure}
% % \centering
% \includegraphics[width=0.7\textwidth]{sentence}
% \caption{A check-in sequence of \textit{feature words}.}
% \label{fig:sentence}
% \end{figure} 

\textbf{Users/Neighborhoods.} Depending on whether the task is user-centric (location recommendation) or area-centric (urban functional zone study and crime prediction), a user $u$ or a neighborhood $n$ is taken as the context from which check-in sequences are extracted. 

Given \textit{feature word} sequences and \textit{location word} sequences, our model learns embedding vectors of \textit{feature words} and \textit{location words} independently in two semantic spaces. Based on these intermediate representations we calculate the vectors corresponding to check-ins, users, venues, and neighborhoods. This aggregation process will be elaborated in following sections.

% Our model learns embedding vectors of \textit{feature words} and \textit{location words} independently in two semantic spaces given the check-in sequences. Then we can calculate representations for locations, neighborhoods, and users. 

\subsection{Embedding Model}\label{sec:Emodel}   

% \stepcounter{figure}
\begin{figure}
% \centering
\includegraphics[width=0.7\textwidth]{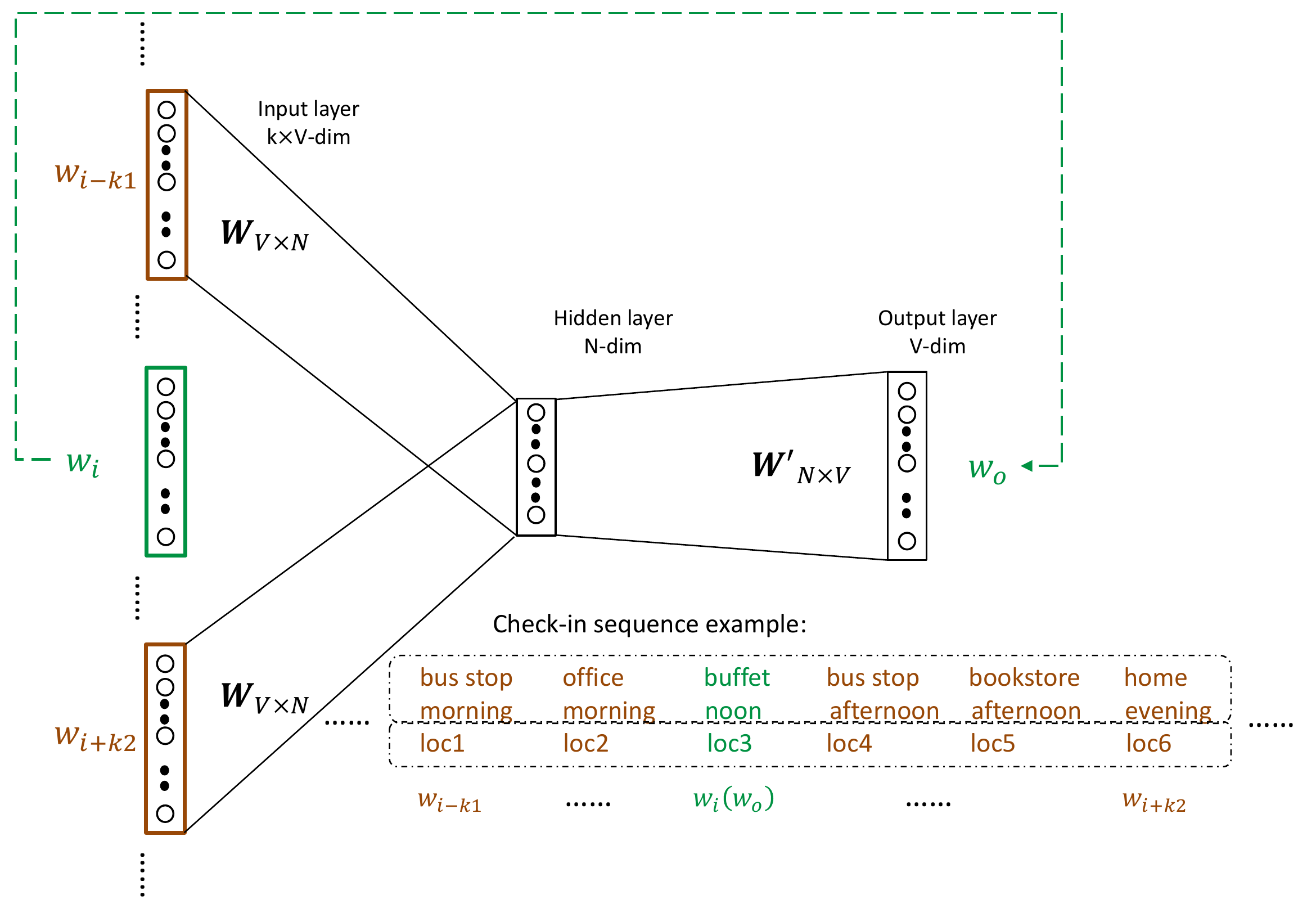}
\caption{The embedding training framework. Each target word (highlighted in green) takes turns to be predicted from its context words (highlighted in brown).}
\label{fig:NN}
\end{figure}   

Figure~\ref{fig:NN} illustrates the neural network (NN) training framework. In the spirit of~\cite{mikolov2013distributed}, each target check-in word (\textit{feature word} or \textit{location word}) $w_i$($w_o$) is predicted from preceding and following check-ins in a sliding window from $w_{i-k1}$ to $w_{i+k2}$. Where $k1$ and $k2$ are adjustable and $k = k1 + k2$ is the overall context window size. All check-in words are initialized with one-hot encoded vectors, which means for a given word, only one out of $V$ vector components will be 1 while all others are 0. When the training process is finished, the row of weight matrix $\mathbf{W}_{V \times N}$ from input layer to hidden layer is the vector representation of the corresponding word. 

The training objective is to minimize the loss function
%\begin{displaymath}
\begin{equation}
\label{eqn:eq1}
E = - log\ p(w_o|w_{i-k1},...,w_{i+k2}).
\end{equation}
%\end{displaymath}
        
where $p(w_o|w_{i-k1},...,w_{i+k2})$ is the probability of the target check-in given the context check-ins, which can be formulated as a softmax function

%\begin{displaymath}
\begin{equation}
\label{eqn:eq2}
p(w_o|w_{ik}) = \frac{\exp({\mathbf{v}_{w_{o}}^{\prime}}^T\mathbf{v}_{w_{ik}})}{\sum_{j=1}^{V} \exp({\mathbf{v}_{w_{j}}^{\prime}}^T\mathbf{v}_{w_{ik}})},
\end{equation}
%\end{displaymath}

where
%\begin{displaymath}
\begin{equation}
\label{eqn:eq3}
\mathbf{v}_{w_{ik}} = \frac{1}{k}(\mathbf{v}_{w_{i-k1}}+...+\mathbf{v}_{w_{i+k2}}).
\end{equation}
%\end{displaymath}

In Equation~\ref{eqn:eq2}, $\mathbf{v}_w^{\prime}$s comes from the columns of $\mathbf{W}^{\prime}_{N \times V}$, the weight matrix connecting hidden layer to output layer. Back-propagation is applied during the training process and both hidden-to-output weights ($\mathbf{W}^{\prime}$) and input-to-hidden weights ($\mathbf{W}$) are updated using stochastic gradient descent.

Also, from Equation~\ref{eqn:eq2}, we can see that the learning process involves a traversal of all \textit{feature words} or \textit{location words}, which may jeopardize model efficiency. A typical method to tackle this problem is to employ the hierarchical softmax algorithm as proposed in~\cite{mikolov2013exploiting}. To do so, we construct a Huffman binary tree~\cite{knuth1985dynamic}, in which $V$ vocabulary words are leaf units, and for each of them, there exists a unique path to the root. We only consider the along-path words when calculating the loss function. Our preliminary experiments demonstrate that utilizing hierarchical softmax improves the time efficiency by around 13\%. On the other hand, the location recommendation accuracies after using hierarchical softmax are in most cases comparable with the raw results. The largest loss is approximately 0.7\%.

\textit{Feature words} and \textit{location words} are separately trained via this model, resulting in a feature embedding space and a geographic embedding space. Now we can represent a check-in $c$ by only its \textit{feature word} vector or only its \textit{location word} vector. Instead, to obtain a single joint representation, we follow~\cite{he2016deep} in summing up its \textit{feature word} and \textit{location word} vectors in element-wise fashion. 

\begin{figure}
% \centering
\includegraphics[width=0.7\textwidth]{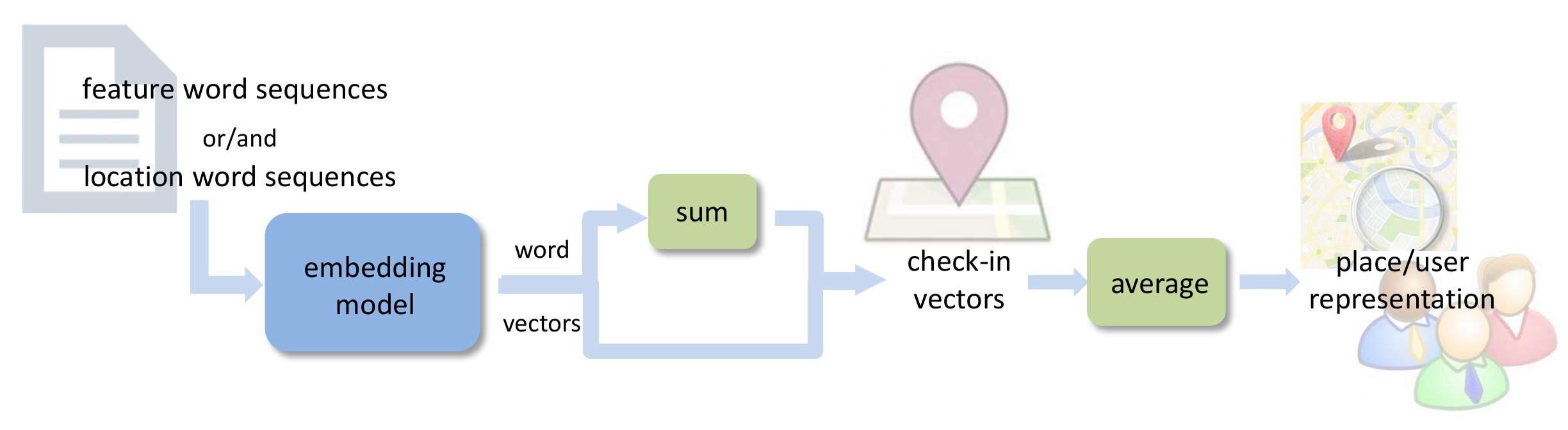}
\caption{Flowchart from importing check-ins to obtaining embedding vectors of places or users.}
\label{fig:flowchart}
\end{figure}

Furthermore, a user $u$ can be represented by the mean of his/her check-in vectors, which also works if we only want to profile their activities in a specific time window. The same approach is applied to annotate a place or a neighborhood. Figure~\ref{fig:flowchart} demonstrates the entire workflow.

% \stepcounter{figure}

% To conclude the \textit{word2vec} analogy that motivates our network training process, we can conceptually think of  \textit{feature words} and \textit{location words} of check-in records as ``\textit{words}'', check-in sequences as ``\textit{sentences}'', and check-in profiles of users or neighborhoods as ``\textit{documents}''.

\subsection{Recommendation Algorithm}\label{sec:Ralgo}
Our recommendation algorithm is based on the user-location cosine similarity in the newly established embedding space. Recall that in the literature review, we mentioned how both temporal and geographic elements play important roles in recommendation tasks. Therefore, we utilize both \textit{feature word} vectors ($\mathbf{v}_{fw}$) and \textit{location word} vectors ($\mathbf{v}_{gw}$) to make recommendations. In this case, a check-in ($\mathbf{v}_c$) is represented by the element-wise summation of these two vectors
%\begin{displaymath}
\begin{equation}
\label{eqn:eq4}
\mathbf{v}_c = \mathbf{v}_{fw} + \mathbf{v}_{gw}.
\end{equation}

\begin{figure}
% \centering
\includegraphics[width=0.5\textwidth]{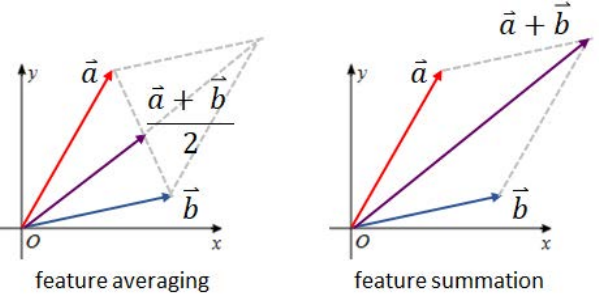}
\caption{Both averaging and summation fuse constituent vectors into a single final representation. Without the re-scaling constant, summation better preserves the information carried by the original feature vectors.}
\label{fig:average_summation}
\end{figure}

Such element-wise summation of feature vectors from different spaces has been successfully implemented in the field of computer vision~\cite{he2016deep} when training deep neural networks. As illustrated in Figure~\ref{fig:average_summation}, similar to averaging, summation fuses features but without applying a re-scaling constant, which better preserves the information carried by the original feature vectors. In our preliminary experiments, we examined the location recommendation accuracy utilizing the check-in vectors calculated by summation, averaging, and concatenation respectively. The results clearly indicate that summation-based vectors outperform the alternatives by $4-5\%$ absolute performance.

% \stepcounter{figure}

On top of these summation-based vector representations, we profile locations and users. Remember that functional roles are defined by social network venue categories. For the sake of clarity, we will refer to ``venue category'' in place of ``functional role'' in the remainder of this section.

\textbf{Location Profile.} Although two locations may belong to the same venue category, they can still be differentiated if they are usually visited in different timeslots. A location $l$ can thus be represented as $\mathbf{v}_l$ by averaging all user check-ins $\mathbf{v}_c$ issued there:
%\begin{displaymath}
\begin{equation}
\label{eqn:eq5}
\mathbf{v}_l = \frac{1}{M} \sum_{m=1}^{M} \mathbf{v}_{c_m},
\end{equation}
%\end{displaymath}
where $M$ is the total count of check-ins originating from location $l$.

\begin{table}
 %\scriptsize
  \caption{Top 3 most visited venue categories in different timeslots.}
  \label{tab:preference}
  \begin{tabular}{lc}
    \toprule
     Timeslot & Top 3 popular venue categories \\
    \midrule
    Morning & \textit{Professional\&Other places}(25.9\%), \textit{Food}(23.4\%), \textit{Travel\&Transport}(21.9\%)
    \\
    Noon & \textit{Food}(40.2\%), \textit{Professional\&Other Places}(17.1\%), \textit{Shop\&Service}(11.6\%) \\
    Afternoon & \textit{Food}(27.9\%), \textit{Travel\&Transport}(15.2\%), \textit{Shop\&Service}(14.1\%) \\
    Evening & \textit{Food}(30.3\%), \textit{Nightlife Spots}(19.5\%), \textit{Arts\&Entertainment}(14.3\%)\\
    Night & \textit{Nightlife Spots}(35.2\%), \textit{Food}(22.6\%), \textit{Travel\&Transport}(11.7\%) \\
    WeekendMorning & \textit{Food}(26.8\%), \textit{Outdoors\&Recreation}(21.4\%), \textit{Travel\&Transport}(18.9\%) \\
    WeekendNoon & \textit{Food}(36.1\%), \textit{Outdoors\&Recreation}(16.0\%), \textit{Shop\&Service}(14.3\%) \\
    WeekendAfternoon & \textit{Food}(30.1\%), \textit{Outdoors\&Recreation}(16.7\%), \textit{Shop\&Service}(15.8\%)\\
    WeekendEvening & \textit{Food}(33.6\%), \textit{Nightlife Spots}(16.6\%), \textit{Arts\&Entertainment}(14.0\%) \\
    WeekendNight & \textit{Nightlife Spots}(49.4\%), \textit{Food}(21.1\%), \textit{Arts\&Entertainment}(8.4\%) \\
    \bottomrule
  \end{tabular}
\end{table}

\textbf{User Profile.} In a preliminary study, we confirm Li \textit{et al.}'s hypothesis of check-in distributions differing across timeslots~\cite{li2011tweet} (see Table~\ref{tab:preference}). Correspondingly, and following intuition, frequently visited places vary according to time-of-day. 
% In different timeslots, users show different check-in preferences in terms of venue categories and geographic areas.
Inspired by this observation, we calculate 10 profiles for each user corresponding to different timeslots. In each timeslot $t$, we represent a user $u$ as $\mathbf{v}_{u_t}$ by averaging all his/her check-ins $\mathbf{v}_{c_t}$ in this timeslot and calculate a user coordinate centroid ($coordinate_{u_t}$) from those check-in locations ($coordinate_{c_t}$). 
%\begin{displaymath}
\begin{align}
\mathbf{v}_{u_t} &= \frac{1}{N}\sum_{n=1}^{N} \mathbf{v}_{c_{t,n}} \\
coordinate_{u_t} &= \frac{1}{N} \sum_{n=1}^{N} coordinate_{c_{t,n}}
\end{align}
%\end{displaymath}
where $N$ is the total count of check-ins from user $u$ in timeslot $t$.

Next, we calculate two \textit{cosine similarities} for users in \textit{each} timeslot. The first one, \textit{user-activity similarity} $S_{u-a}$, relates the user vector $\mathbf{v}_{u_t}$ to every check-in vector of this user ($\mathbf{v}_{c,u}$). This similarity indicates the time-wise user-preferred venue categories. The second one, \textit{user-location similarity} $S_{u-l}$, relates the user vector $\mathbf{v}_{u_t}$ to every location vector $\mathbf{v}_{l}$. This similarity indicates the user-preferred locations in each venue category.

During the recommendation stage, given a timeslot $t$, we first select the $C$ most favored venue categories of the user based on the \textit{user-activity similarity} $S_{u-a}$ and list these categories in descending order $f_1,...,f_C$. Now, we focus on unique locations within selected categories. We mark the aforementioned \textit{user-location similarity} $S_{u-l}$ as $S_{u-l,original}$. On basis of it, for each location, we calculate its distance $dist$ to the user coordinate centroid in this timeslot ($coordinate_{u_t}$). Considering the category preference order and location-to-centroid distances, we introduce two exponential decay factors, \textit{category decay CD} and \textit{spatial decay SD}, modeling the likelihood of a user straying from their usual categorical and spatial patterns. 
%\begin{displaymath}
\begin{align}
CD &= a_1 \times \exp(- a_2 \times f_c) \\
SD &= b_1 \times \exp(- b_2 \times dist)
\end{align}
%\end{displaymath}

Where, $a_1, a_2, b_1, b_2 \in \mathbb{R}$, $f_c \in \{0,1,...,C-1\}$. Inspired by an intuitive heuristic which works widely in practice~\cite{koren2010collaborative, heylighen2002hebbian}, we multiply the original user-location similarity with these two decay factors to calculate the final user-location similarity ($S_{u-l,final}$) as
%\begin{displaymath}
\begin{equation}
\label{eqn:eq6}
S_{u-l,final} = S_{u-l,original} \times CD \times SD
\end{equation}
%\end{displaymath}

Afterwards, we sort all locations belonging to these $C$ categories in descending order of $S_{u-l,final}$ and make recommendations from the top. We refer to this algorithm as the \textit{\textbf{S}patial-\textbf{T}emporal \textbf{E}mbedding \textbf{S}imilarity algorithm} (STES) and use the acronym STES in the rest of the paper.

\section{Experiments \& Evaluation}\label{sec:experiments}

In this section, we begin by introducing the experimental data set and data pre-processing details, then we elaborate on the various experiments and evaluate the results. 

% Our empirical investigation is driven by the following research questions:

% \textbf{RQ1.} How well does the embedding model differentiate locations and users along temporal, geographic, and functional aspects?

% \textbf{RQ2.} How does the STES algorithm perform in location recommendation compared with state-of-the-art recommendation methods?

% \textbf{RQ3.} How to define and visualize urban functional zones using the embedding model?

% \textbf{RQ4.} How well can we predict typical urban characteristics based on the embedding model?

% \textbf{RQ5.} With what generalization error can an embedding model trained in one city be transferred to other cities?

\subsection{Data set}

As described in previous sections, the data set is required to contain check-in time, location, and the functional role of the visited venue. A robust and popular method to define venue functional roles is to leverage the Foursquare hierarchy of venue categories. The Foursquare venue category tree has four hierarchical levels. In our work, we utilize the second level categories containing 422 classes such as \textit{American Restaurant}, \textit{Bar}, and \textit{Metro Station}. This is motivated by two reasons. First, there exist only 10 top level labels, which are too coarsely divided to differentiate places. Secondly, third and fourth-level categories are too specific to cover all of the venues. In contrast, second-level categories achieve the best sparsity-specificity trade-off. 

Among data sets containing Foursquare check-ins, we select a publicly available one from~\cite{cheng2011exploring} for three reasons. 
\begin{itemize}
\item \textbf{Space and time span.} This data set contains globally collected check-ins across 11 months from Feb 25. 2010 to Jan 20. 2011, providing over 12 million Foursquare check-in records with the global spread that we require for our experiment about model generalization (RQ5). 
\item \textbf{Sufficient information.} In this data set, each raw check-in entry involves user ID, location coordinates, time, venue ID, and source URL. Although the venue category is not originally included, it can be crawled via Foursquare's venue search API\footnote[3]{https://developer.foursquare.com/docs/venues/venues} using the source URL.
% we leverage the source URL to crawl the category via Foursquare venue search API\footnote[3]{https://developer.foursquare.com/docs/venues/venues}.  
\item \textbf{Comparability.} This data set has been utilized in a variety of relevant works, including location recommendation~\cite{hu2013spatio, hu2013spatial, xie2016learning}, urban activity pattern understanding~\cite{hasan2013understanding, hasan2014urban}, location-based services with privacy awareness~\cite{wernke2012pshare, riboni2013platform}, etc. 
\end{itemize}

Another two Foursquare data sets from~\cite{li2015rank} and \cite{zhao2016gt} are also leveraged in relevant pieces of research. However, the former only contains check-ins in Singapore while the latter does not include any venue category. Therefore, we exclude them from our study. 

% We rely on a publicly available data set from~\cite{cheng2011exploring} which contains check-ins from Foursquare, Twitter, Gowalla, and other location based social networks. This data set is global and covers 11 months from Mar.\ 2010 to Jan.\ 2011. Most of the check-ins are distributed in the U.S.\ and we choose New York City (NYC) as the representative city for our study.

% To define functional roles of venues, we use the second level Foursquare venue categories, containing 422 classes such as \textit{American Restaurant}, \textit{Bar}, and \textit{Metro Station}. This level achieves the best sparsity-specificity trade-off in comparison with the top level which is too coarse and the third or the fourth levels which are too specific to cover all venues.  

A common issue in check-in data streams are repeated check-ins at the same venue in an artificially short time window during which users remain in an unchanged location and activity~\cite{li2016probabilistic}. This appears reasonable as check-ins are often posted in a casual way in which people share real-time affairs and moods. This effect is especially common in recreational and culinary activities. For instance, a user may check in for several times during one meal with friends, each time posting about a newly served dish or commenting on the food. To model activity sequences most reliably, we delete such repeated check-ins from an individual staying in an unchanged activity and retain only the first check-in at this location. To reduce noise, we further remove both users and locations with less than 10 posts. After this pre-processing, for the example of New York City (NYC), our dataset contains 225,782 check-ins by 6,442 users at 7,453 locations. 

To define urban ``neighborhoods'' for our area-centric tasks \textit{urban functional zone study} and \textit{crime prediction}, we utilize official 2010 Census Block Group (CBG) polygons\footnote[4]{http://data.beta.nyc/dataset/2010-census-block-groups-polygons}, matching the time period of the check-in data collection. A CBG may contain several Census Blocks (CBs), which are the smallest geographic areas that the U.S.\ Census Bureau uses to collect and tabulate census data. These polygons represent the most natural segmentation of a city, given that their boundaries are defined by physical streets, railroad tracks, bodies of water as well as invisible town limits, property lines, and imaginary extensions of streets. In NYC, there are 6,493 CBGs, 1,720 of which are populated by our denoised check-ins.

\begin{figure}
% \centering
\subfigure[functionality: mean cosine similarities]{
\label{fig:cate_cosine}
\includegraphics[width=0.45\textwidth]{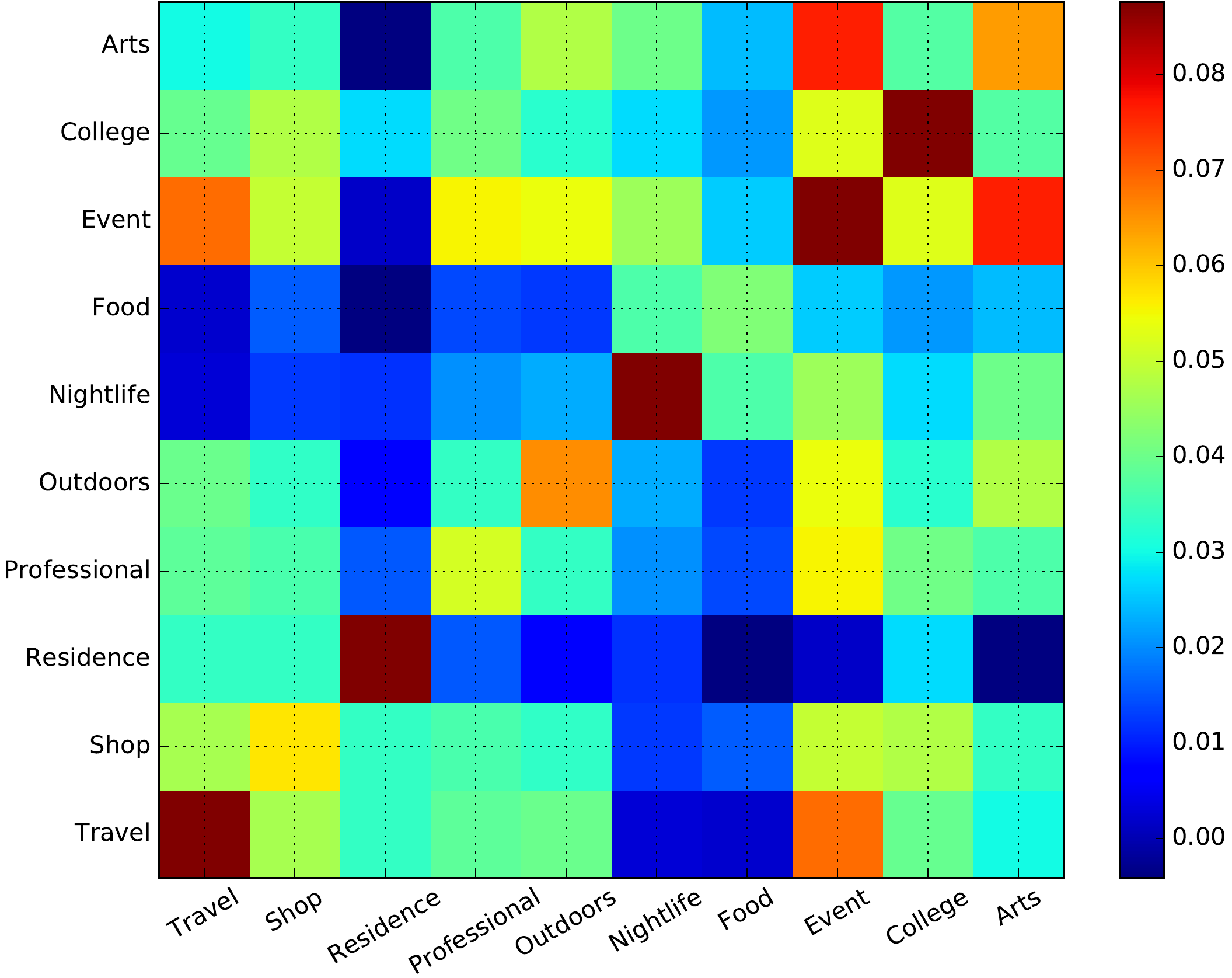}}
\subfigure[functionality: mean euclidean distances]{
\label{fig:cate_euclidean}
\includegraphics[width=0.45\textwidth]{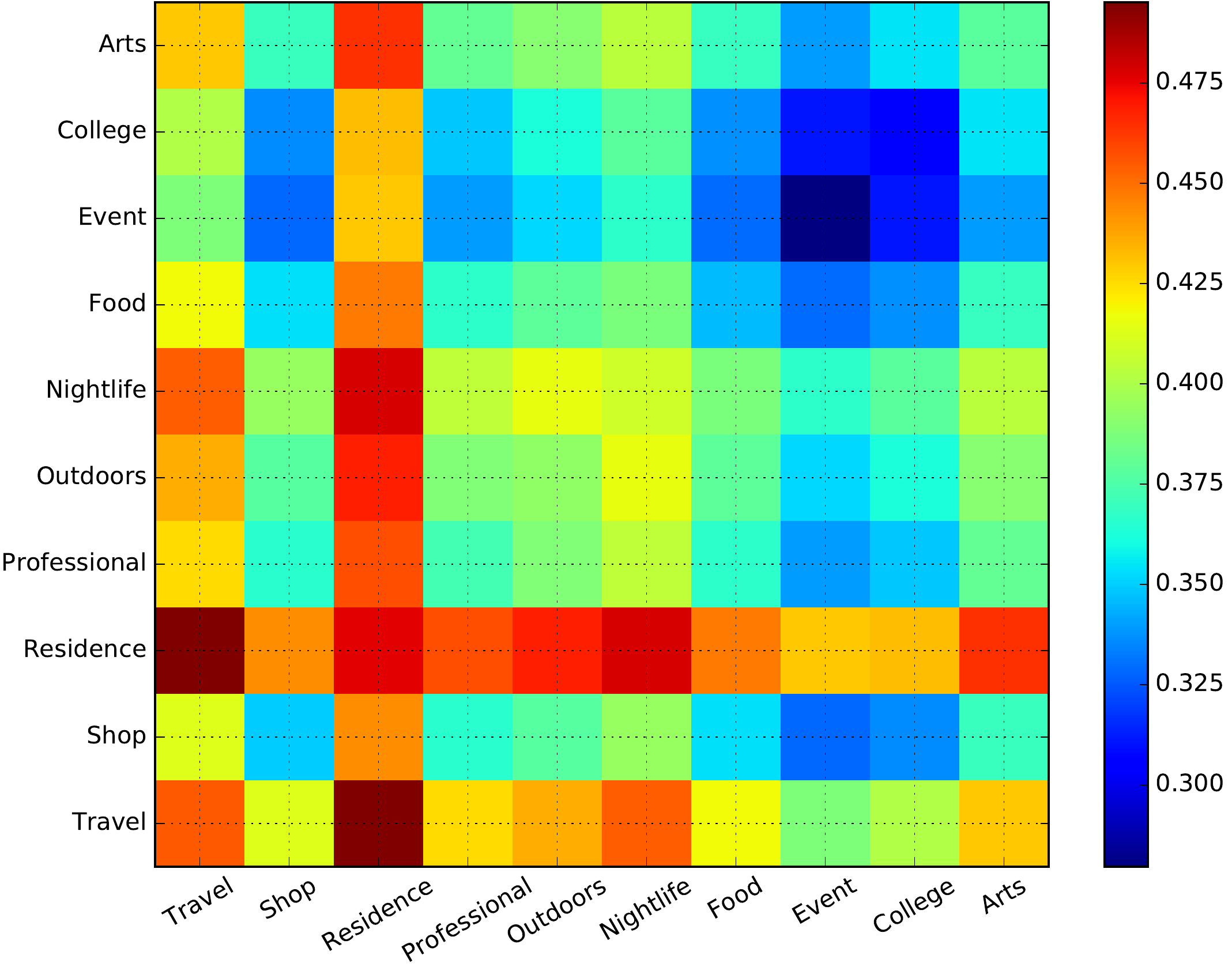}}
\subfigure[time: mean cosine similarities]{
\label{fig:time_cosine}
\includegraphics[width=0.45\textwidth]{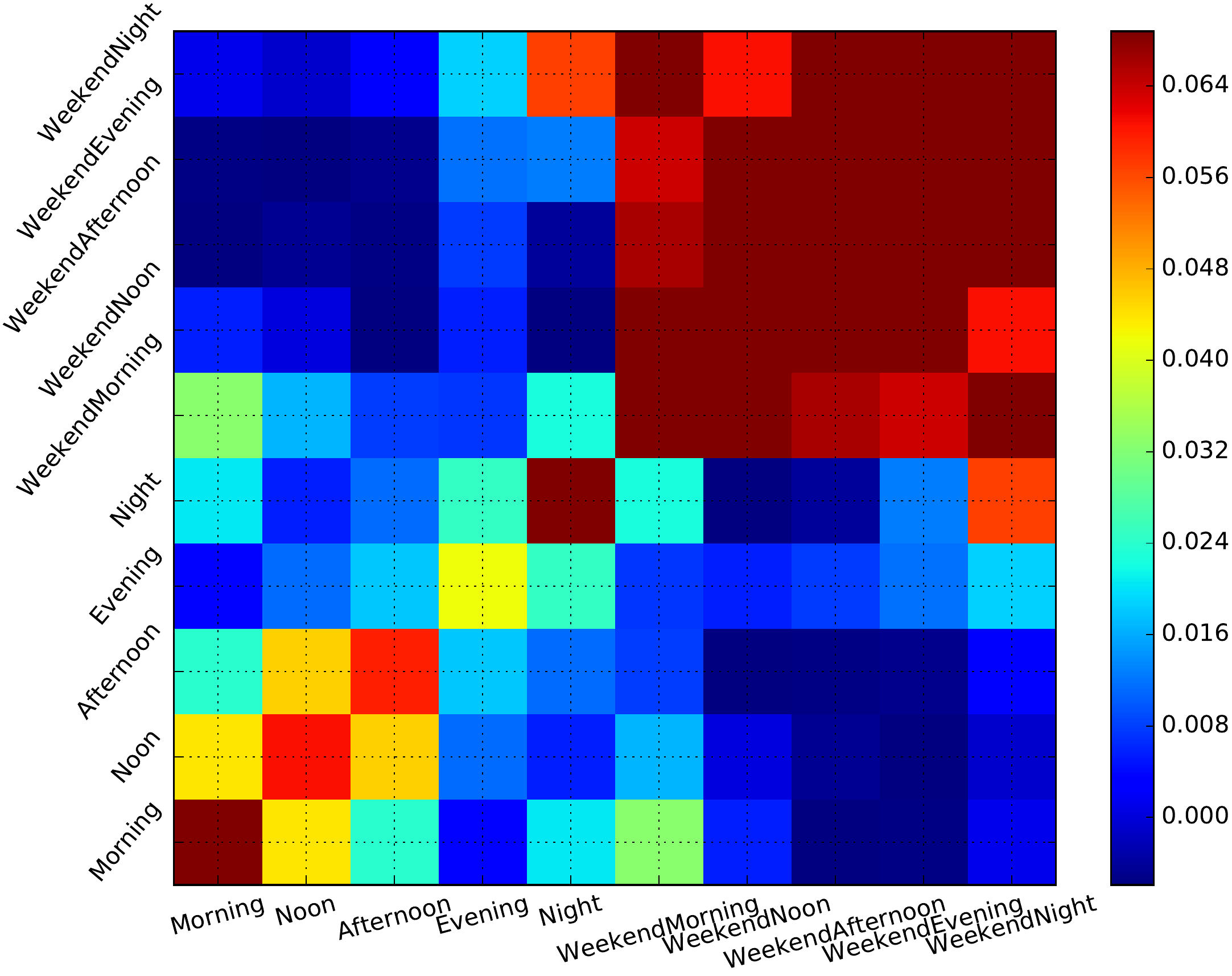}}
\subfigure[time: mean euclidean distances]{
\label{fig:time_euclidean}
\includegraphics[width=0.45\textwidth]{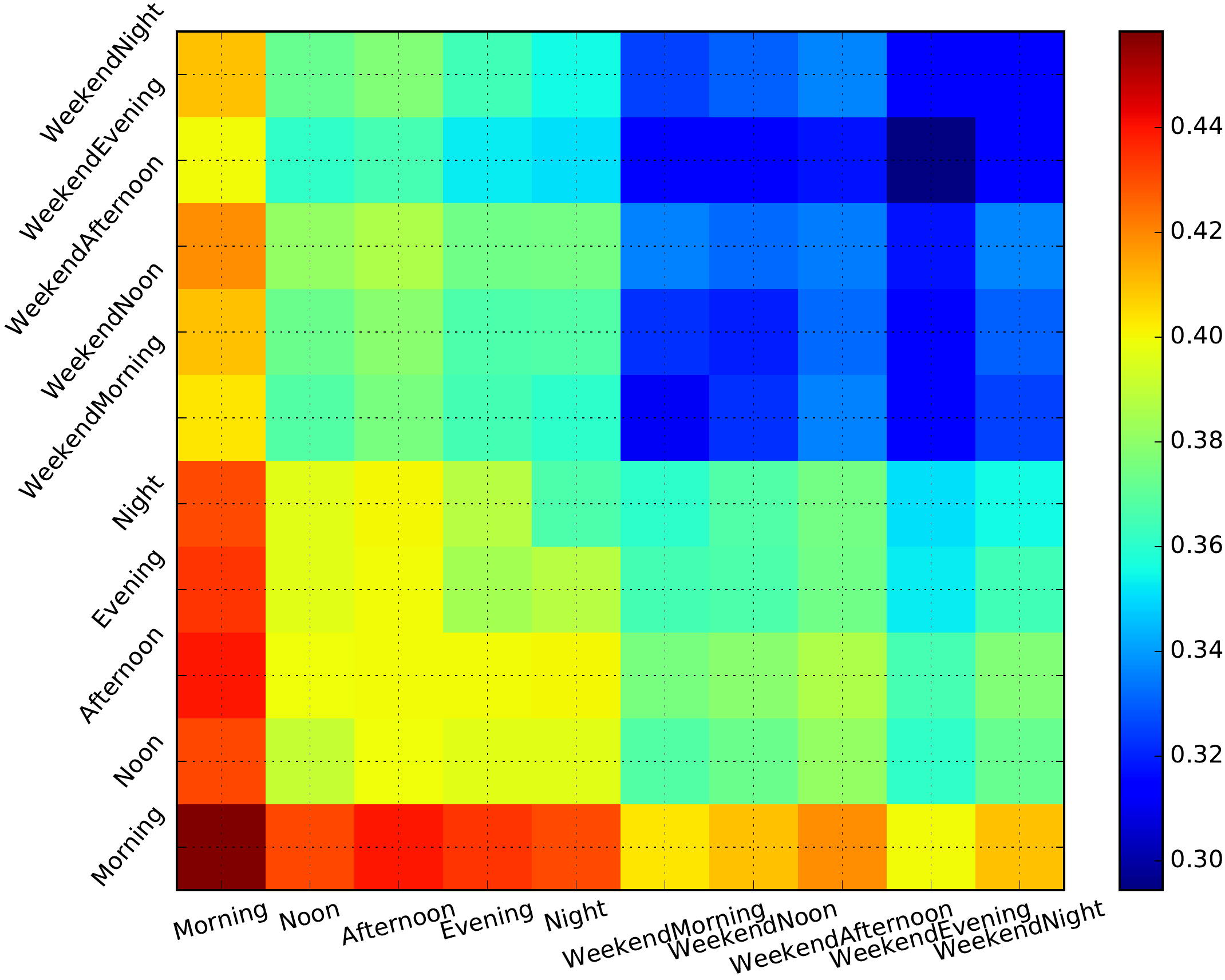}}
\caption{Heatmaps of mean cosine similarities and Euclidean distances for \text{feature word} embeddings. We normalize the Euclidean distances by the distance range.}
\label{fig:venuetime}
\end{figure}

% \begin{figure}
% % \centering
% \subfigure[Cosine similarity of location pairs (y-axis) with increasing geographic distance (x-axis)]{
% \label{fig:loc_cosine}
% \includegraphics[width=0.45\textwidth]{loc_cosine}}
% \subfigure[Euclidean distance of location pairs (y-axis) with increasing geographic distance (x-axis)]{
% \label{fig:loc_euclidean}
% \includegraphics[width=0.45\textwidth]{loc_euclidean}}
% \caption{Cosine similarity and Euclidean distance of location pairs as a function of geographic distance (in $km$). As before, we normalize Euclidean distances by distance range.}
% \label{fig:locline}
% \end{figure}

\subsection{Qualitative Analysis of Embedding Vectors}
Our embedding vectors are designed to retain key characteristics of user activities and urban venues so that users and places are differentiable. We set the latent embedding dimension to 200 and obtain embedding vectors for all of the \textit{feature words} and \textit{location words} in NYC. 

To get a qualitative impression of the resulting embeddings, we first examine their overall cosine similarities and Euclidean distances. The cosine metric evaluates the similarity by normalizing vectors and measuring the in-between angle, and the Euclidean distance demonstrates the magnitude of difference between two vectors. 

We explore \textit{feature word} embedding vectors from both functionality and time perspectives. We first calculate the cosine similarity and Euclidean distance for each pair of embedding vectors, then we compute the mean similarity and distance values among the timeslots and top-level venue categories, respectively. We demonstrate the results in the form of heatmaps in Figure~\ref{fig:venuetime}, where four significant tendencies can be observed. 

In Figure~\ref{fig:cate_cosine}, we can see that intra-category embedding vectors show the highest mean cosine similarities. The single exception \textit{Arts} lists \textit{Arts}-\textit{Arts} similarity as second to the \textit{Arts}-\textit{Event} pair. Considering that \textit{Event} mainly involves places for sport, musical, and arts activities, this result appears reasonable. Categories including similar or overlapping venues also have good inter-class similarities, such as \textit{Food}-\textit{Nightlife}, \textit{College}-\textit{Event}, and \textit{Outdoors}-\textit{Event}.

Turning to Figure~\ref{fig:cate_euclidean}, the mean intra-category Euclidean distances of \textit{College} and \textit{Event} are smaller than inter-category distances, which indicates that these two categories have the most compact embedding vector clouds. \textit{Arts}, \textit{Food}, \textit{Nightlife}, \textit{Outdoors}, \textit{Professional}, and \textit{Shop} also have moderate intra- and inter-class Euclidean distances. \textit{Residence} and \textit{Travel} have the largest intra- and inter-category Euclidean distances, which implies that embedding vectors belonging to these two categories are widely distributed in the embedding space. The overall scenario is akin to a geographic area in reality: \textit{Travel} and \textit{Residence} spots are usually spread over the city, while places with highly specific functions such as \textit{College} are more concentrated in a small district. Venues related to \textit{Food}, \textit{Nightlife}, \textit{Shop}, and \textit{Professional} (mainly including places for work, public services, and medical treatments) are widely spread but usually there also exist specific districts such as a central business districts (CBD), mainly reserved for these activities.

Figure~\ref{fig:time_cosine} shows that check-ins on weekends are more similar to each other while weekday-weekend similarities are less significant. This indicates people's different activity patterns on weekdays and weekends. For each timeslot on weekdays, the highest mean cosine similarity comes from the intra-timeslot case. We can also see that the difference between working hour activities and after-work life is delivered by the similarities within daytime (\textit{Morning}, \textit{Noon}, and \textit{Afternoon}) and nighttime (\textit{Evening} and \textit{Night}) embedding vectors. It also demonstrates the continuity of time under our model. For instance, \textit{Morning} vectors are similar to \textit{Noon} vectors, and \textit{Noon} vectors are similar to \textit{Afternoon} vectors. However, \textit{Morning}-\textit{Afternoon} similarity is much less significant.

Figure~\ref{fig:time_euclidean} shows that vector clusters for weekend timeslots are more compact compared to weekday ones. As intra-timeslot vectors involve all of the activities taking place in that period of time, this distance heatmap underlines the fact that in contrast to the weekend life which mainly involves leisure and recreation, diversity is much greater on weekdays where professional and educational activities are included as well. 

% \begin{figure}
% % \centering
% \subfigure[functionality: mean cosine similarities]{
% \label{fig:cate_cosine}
% \includegraphics[width=0.45\textwidth]{cate_cosine}}
% \subfigure[functionality: mean euclidean distances]{
% \label{fig:cate_euclidean}
% \includegraphics[width=0.45\textwidth]{cate_euclidean}}
% \subfigure[time: mean cosine similarities]{
% \label{fig:time_cosine}
% \includegraphics[width=0.45\textwidth]{time_cosine}}
% \subfigure[time: mean euclidean distances]{
% \label{fig:time_euclidean}
% \includegraphics[width=0.45\textwidth]{time_euclidean}}
% \caption{Heatmaps of mean cosine similarities and Euclidean distances for \text{feature word} embeddings. We normalize the Euclidean distances by the distance range.}
% \label{fig:venuetime}
% \end{figure}

\begin{figure}
% \centering
\subfigure[Cosine similarity of location pairs (y-axis) with increasing geographic distance (x-axis)]{
\label{fig:loc_cosine}
\includegraphics[width=0.45\textwidth]{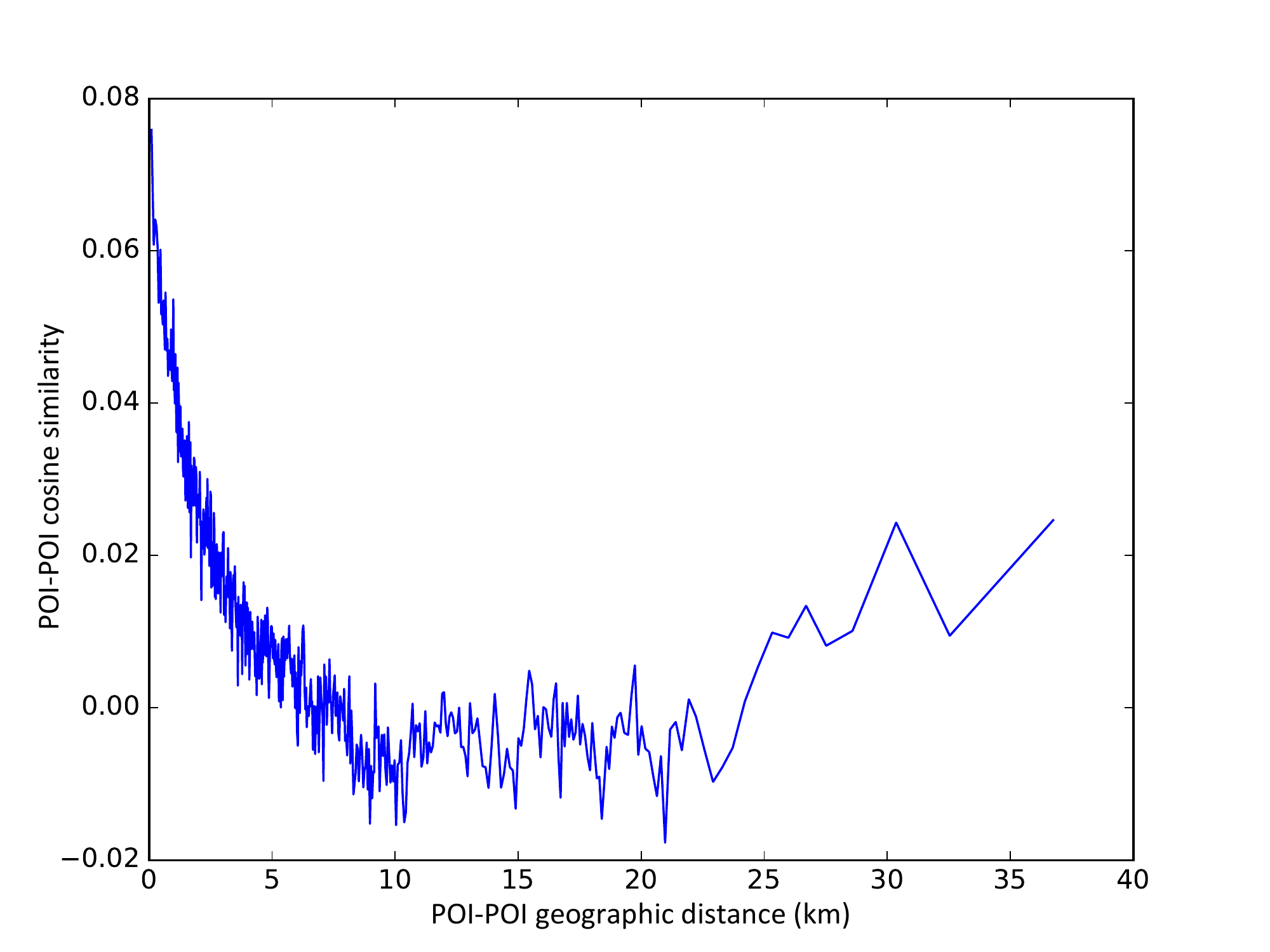}}
\subfigure[Euclidean distance of location pairs (y-axis) with increasing geographic distance (x-axis)]{
\label{fig:loc_euclidean}
\includegraphics[width=0.45\textwidth]{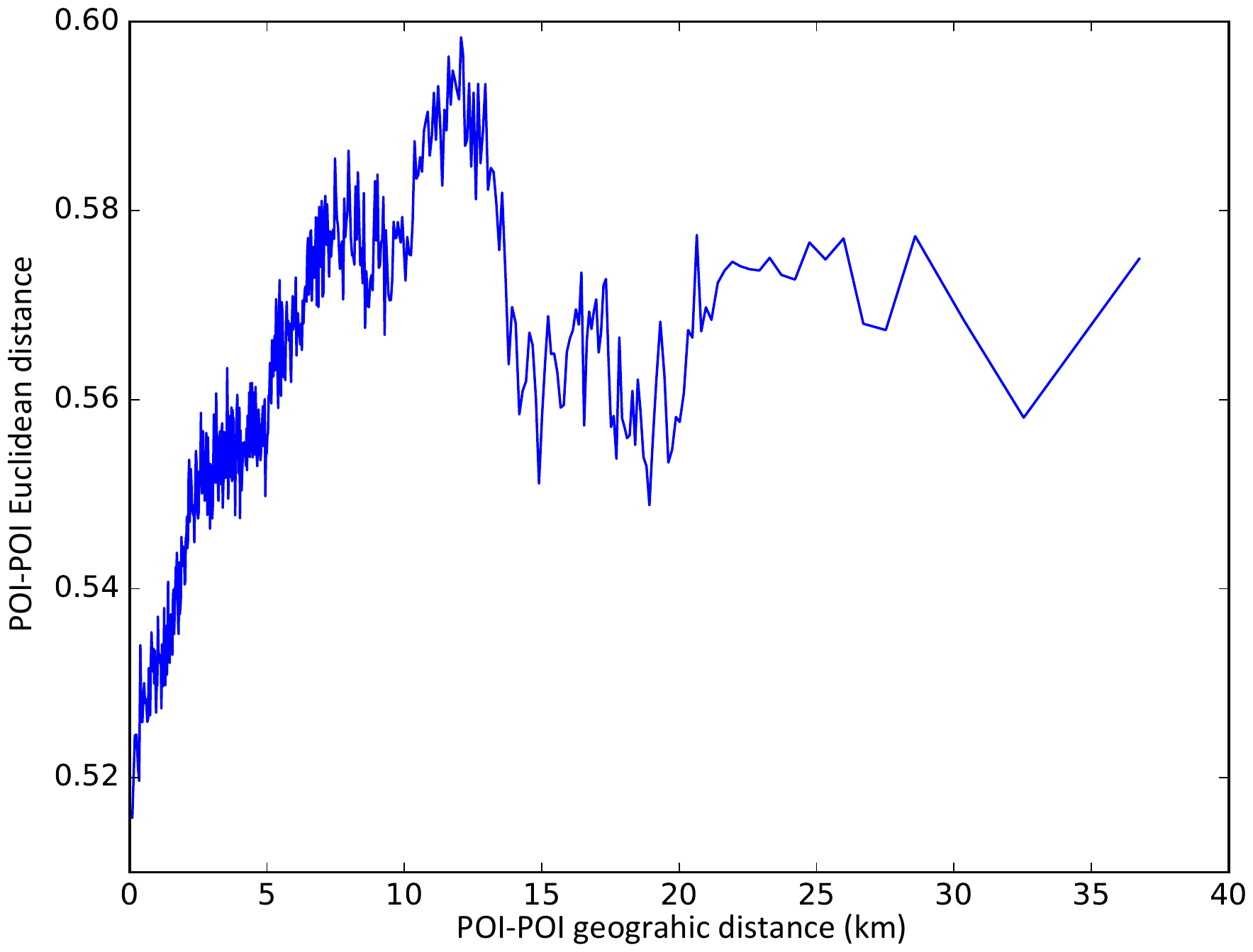}}
\caption{Cosine similarity and Euclidean distance of location pairs as a function of geographic distance (in $km$). As before, we normalize Euclidean distances by distance range.}
\label{fig:locline}
\end{figure}

Now we examine the \textit{location word} embedding vectors. As for \textit{feature word} vectors, we first calculate the cosine similarity and Euclidean distance for each pair of locations. We also calculate their geographic distances in the physical world. We sort the resulting triples (geographic distance, cosine similarity, Euclidean distance) according to increasing geographic distances.

To display the results in a legible and informative manner, we divide raw sequences into short segments of length 500. We calculate their mean distances and similarities per segment and plot the results in Figure~\ref{fig:locline}. 

% \begin{figure}
% % \centering
% \subfigure[Cosine similarity of location pairs (y-axis) with increasing geographic distance (x-axis)]{
% \label{fig:loc_cosine}
% \includegraphics[width=0.45\textwidth]{loc_cosine}}
% \subfigure[Euclidean distance of location pairs (y-axis) with increasing geographic distance (x-axis)]{
% \label{fig:loc_euclidean}
% \includegraphics[width=0.45\textwidth]{loc_euclidean}}
% \caption{Cosine similarity and Euclidean distance of location pairs as a function of geographic distance (in $km$). As before, we normalize Euclidean distances by distance range.}
% \label{fig:locline}
% \end{figure}

It can be seen that within approximately 10$km$, as the geographic distance between locations increases, there is a decreasing tendency of the cosine similarity and a reverse trend of the Euclidean distance between their embeddings. Beyond the 10$km$ point, both curves demonstrate fluctuation and rebounding trends. Such variations reflect the setup of real urban spaces where districts of similar functionalities are distributed periodically along geographic distances.  

To further verify this relationship, we measure the correlation between geographic distance and cosine similarity/Euclidean distance using Pearson correlation coefficients and Spearman rank-order coefficients respectively. Both coefficients range between -1 and 1 with 0 implying no correlation.

We begin by globally measuring correlations for all POI-POI pairs and note a moderate-to-strong connection between geographic distance and cosine similarity (Pearson: -0.576, Spearman: -0.763). The correlation between geographic distance and vector space Euclidean distance is very similar (Pearson: 0.555, Spearman:0.788).

As observed earlier, in Figure~\ref{fig:locline}, the connection between physical and vector space distances seems more pronounced within a range of 10$km$. To account for this fact, we further measure correlation coefficients for those POI-POI pairs whose geographic distance does not exceed 10$km$. In this setting, we note a near perfect correlation between physical distance and cosine similarity (Pearson: -0.851, Spearman: -0.934) as well as Euclidean distance (Pearson: 0.933, Spearman: 0.953).

In all of the above cases, the accompanying p-values are much smaller than $0.001$, suggesting that the observations are stable.

% Most of the geographic distances are within 15$km$. As the geographic distance increases, we can see the decreasing tendency of the cosine similarity and the reverse trend of the euclidean distance. 

\begin{figure}
% \centering
\subfigure[\textit{feature word} embeddings]{\label{fig:feature word}
\includegraphics[width=0.45\textwidth]{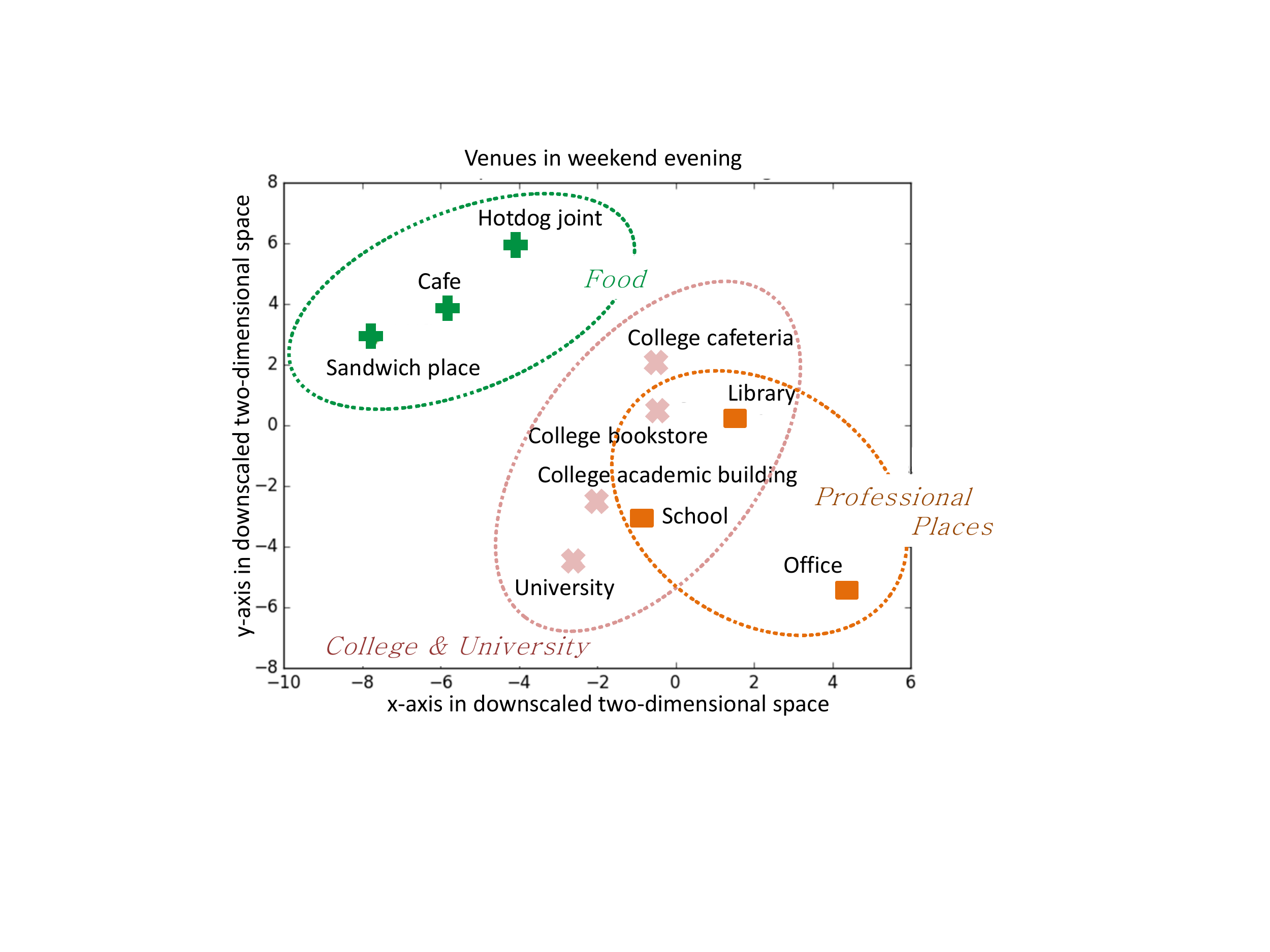}}
\subfigure[\textit{location word} embeddings]{
\label{fig:geographic word}
\includegraphics[width=0.45\textwidth]{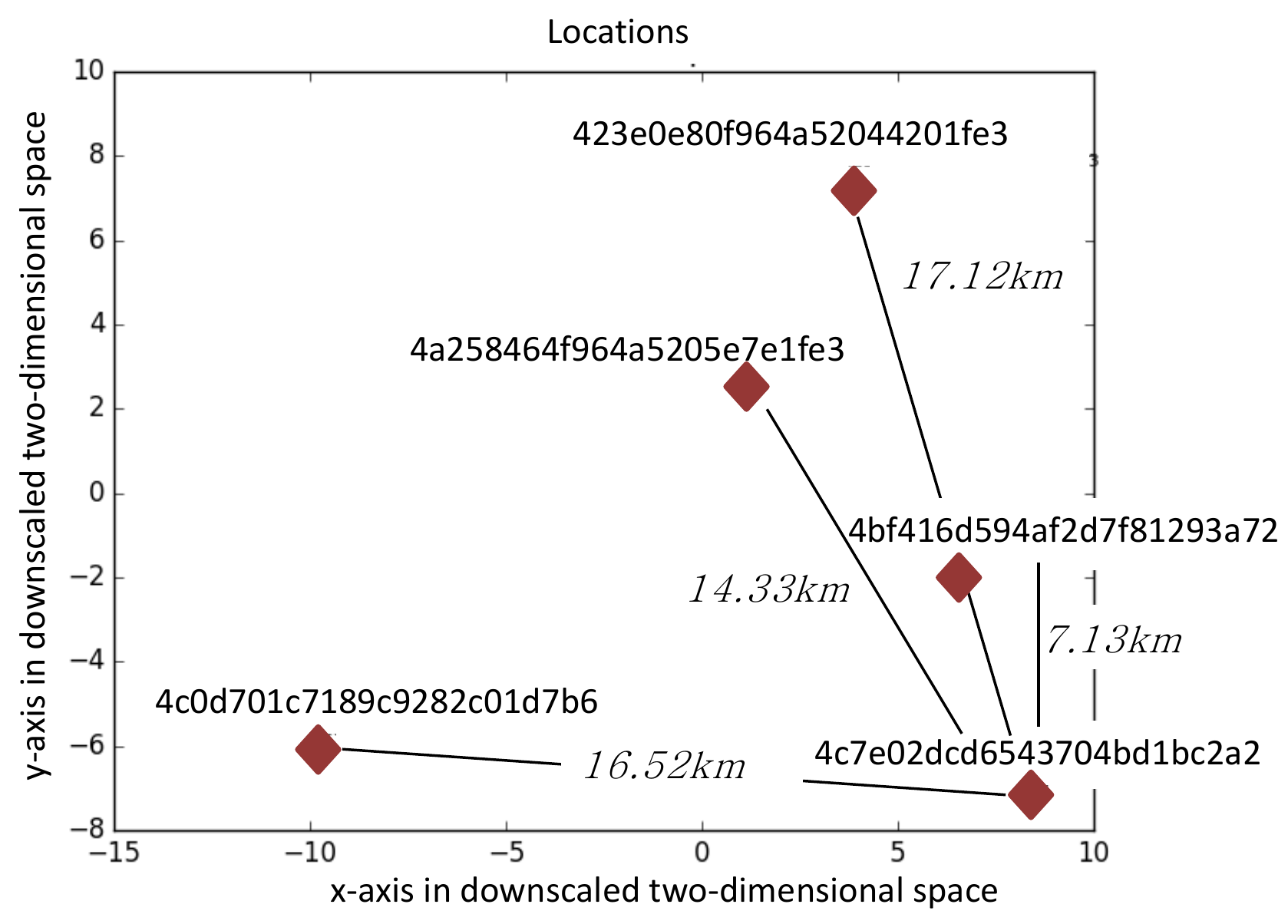}}
\caption{\textit{Feature word} and \textit{location word} example vectors in their semantic spaces.}
\label{fig:word example}
\end{figure}

To more tangibly and intuitively showcase the trained embedding vectors in their semantic spaces, we project and plot the original 200-dimensional vectors into a two-dimensional space. We experimented with various algorithms including isometric mapping (Isomap), multi-dimensional scaling (MDS), t-distributed stochastic neighbor embedding (t-SNE), and finally arrived at MDS as implemented in Python's scikit-learn package\footnote{http://scikit-learn.org/stable/modules/generated/sklearn.manifold.MDS.html} as it produces the most informative visualizations. Examples of these projected vectors are depicted in Figure~\ref{fig:word example}.

% \stepcounter{figure}

% As expected, \textit{feature word} vectors and \textit{location word} vectors are embedded with functional, temporal, and geographic similarities. 
Figure~\ref{fig:feature word} shows some venue embeddings during weekend evenings. \textit{Food} venues, \textit{College\&University} venues, and \textit{Professional} venues are clustered into three groups with the latter two sharing a considerable overlap. This demonstrates the embedding's ability to retain functional correlations that might not have been expressed by discrete venue category labels. More specifically, \textit{School} and \textit{Library} are much closer to \textit{College\&University} places than to other professional places such as \textit{Offices} that do not serve an educational purpose.

\textit{Location word} embeddings reflect geographic proximity among venues. In Figure~\ref{fig:geographic word}, we take location \textit{4c7e02dcd6543704bd1bc2a2} as an example. It can be seen that vector-to-vector Euclidean distances qualitatively correspond to location-to-location geographic distances. 

This section aims to qualitatively answer \textbf{RQ1}. \textit{Feature word} vectors and \textit{location word} vectors are embedded with functional, temporal, and geographic similarities. Annotated with such embedding vectors, places and users are represented in a way that reflects location visiting patterns as well as people's activity preferences. In the following, we will quantitatively show how the embedding vectors can be used for specific problems.

\subsection{Location Recommendation}\label{sub:locRec}

The goal of location recommendation is to predict a list of top-k locations that a specific user may visit given a reference timeslot. For each user, we choose his/her first 80\% check-ins as training data and the remaining 20\% as test data. 

Before comparing with other state-of-the-art recommendation algorithms, we first examine several variants of our proposed method in Section~\ref{sec:methodology}. 

\textbf{Variant 1}: Check-ins are represented only by their \textit{feature words}.

\textbf{Variant 2}: Check-ins are represented only by their \textit{location words}.

\textbf{Variant 3}: Both \textit{feature word} and \textit{location word} embeddings are applied to represent a check-in. However, the spatial decay is calculated based on the distance from the most recent check-in instead of the historic user coordinate centroid.

\textbf{Variant 4}: Only \textit{feature word} embeddings are applied to represent a check-in but we adjust the embedding training process. We utilize the \textit{location word} at the output layer to tune model weights.

\textbf{Variant 5}: Similar to Variant 4, we only utilize the \textit{feature word} embeddings but adjust the training process by adding a subsequent second training round. \textit{Feature words} are leveraged as the output and the input layer of the first and the second training round respectively. \textit{Location words} are utilized as the output layer of the second training round. 

The first three variants are mainly related to the recommendation algorithm described in Section~\ref{sec:Ralgo} and the last two variants change the embedding training process elaborated in Section~\ref{sec:Emodel}.

The latent embedding dimension is set to 200. We individually tune the parameters of each variant to reach optimal performance before comparing them with our proposed STES algorithm in terms of top-k accuracy as utilized in \cite{hu2013spatio}. The top-k accuracy for a test check-in is 1 if the ground-truth location is in the top-k recommendations and 0 otherwise. We demote the metric as acc@k and report the average top-1, top-5, and top-10 accuracies over all test check-ins in Table~\ref{tab:variants}. 

\begin{table}
% \tiny
  \caption{Location recommendation performance evaluation of STES algorithm and its variants.}
  \label{tab:variants}
  \begin{tabular}{lcccccc}
    \toprule
     & \textbf{STES} & Variant 1 & Variant 2 & Variant 3 & Variant 4 & Variant 5 \\
    \midrule
    acc@1 & \textbf{0.105} & 0.041 & 0.036 & 0.092 & 0.055 & 0.057 \\
    acc@5 & \textbf{0.176} & 0.077 & 0.063 & 0.154 & 0.103 & 0.109 \\
    acc@10 & \textbf{0.199} & 0.089 & 0.078 & 0.189 & 0.126 & 0.131 \\
    \bottomrule
  \end{tabular}
\end{table}

From Table~\ref{tab:variants} we can see that the STES algorithm outperforms all its variants. Especially when alltogether removing either \textit{feature} or \textit{location words}, we experience harsh performance losses as compared to the overall model. We will now move on to comparing STES with a representative range of state-of-the-art location recommendation algorithms. 

% For comparison, we include five state-of-the-art location recommendation algorithms as performance baselines.

\textbf{STT}~\cite{hu2013spatio}: This is a topic model based method with consideration of geographic influence and temporal activity patterns. Based on user-specific and time-specific topic distributions, the model selects a check-in topic and recommends a location according to the topic and time-dependent location distributions. 

\textbf{GT-SEER}~\cite{zhao2016gt}: This algorithm is based on neural embedding techniques. To represent geographic influence, it compares the distance between two places with a threshold, and then explicitly defines neighboring places. 
% it defines neighboring places by comparing the distance between two places with threshold. 
It also models the temporal variance into latent location representations during the embedding vector training process.

\textbf{TA-PLR}~\cite{liuexploring2016dada}: Another embedding-based approach. Given check-in sequences, it trains embeddings for location IDs. It also trains latent representation vectors for each time frame and each user. Location recommendation is then based on temporal user-location preference.    

\textbf{Rank-GeoFM}~\cite{li2015rank}: This is a ranking-based factorization method incorporating temporal and geographic influence. Assuming that location preference is relative to the check-in frequency, it fits the users' preference rankings for places to learn the latent factors of users and places. This method is further discussed in~\cite{liu2017experimental}, in which 12 recommendation schemes are evaluated. Based on four widely-used metrics precision, recall, normalized discounted cumulative gain, and mean average precision, the Rank-GeoFM method consistently performs the best among different datasets and user types.

\textbf{LRT}~\cite{gao2013exploring}: This matrix factorization based method measures temporal influence to capture user-location preference and makes recommendations accordingly.

\textbf{GEmodel}~\cite{xie2016learning}: This work introduces a graph-based embedding model. The authors design four bipartite graphs to encode sequential effects, geographical influence, temporal cyclic effects, and semantic effects respectively, and train embedding vectors to represent user, time, region, and POI for user preference ranking calculation. 

In addition to the previously studied top-k \textit{accuracy}, we also compare individual system performance in terms of \textit{precision}, \textit{recall}, and \textit{mean average precision (MAP)}, all of which are common choices in location recommendation evaluation~\cite{zhao2016gt, hu2013spatio, liu2016predicting}. Similar to accuracy, we denote these metrics at top-k recommendation as p@k, r@k, and MAP@k respectively. Their definitions are formulated as follows,

% \begin{equation}
\begin{align}
p@k &= \frac{1}{|U|}\sum_{u\in U}\frac{|L_{gt}\cap L_{rec}|}{k} \\
r@k &= \frac{1}{|U|}\sum_{u\in U}\frac{|L_{gt}\cap L_{rec}|}{|L_{gt}|} \\
%acc@k &= \frac{|L_{gt,all}\cap L_{rec,all}|}{|T|} \\
MAP@k &= \frac{\sum_{t\in T}AveP(t)}{|T|}
\end{align}
% \end{equation}

in which, $U$ represents the user set, $L_{gt}$ and $L_{rec}$ represent the set of ground truth locations and the set of corresponding recommended locations for each user in the test data. 
%$L_{gt,all}$ and $L_{rec,all}$ follow the same meanings but for all users. 
$T$ represents the total test data set, and $AveP(t)$ refers to the average precision for each test case.    

To find the individually best performance of each method, we tune parameters on the training set using cross validation. The best results for each method are listed in Table~\ref{tab:nyc_loc} and correspond to the following parameters:
% We report the performances over all of the recommendation methods in Table~\ref{tab:nyc_loc}. Table~\ref{tab:nyc_loc} lists the best performances of all the methods corresponding to the following parameters, \hlcyan{which were selected from experiments based on changing parameters.}

% \stepcounter{table}
\begin{table}
\tiny
  \caption{Location recommendation performance evaluated by precision, recall, accuracy, and MAP.}
  \label{tab:nyc_loc}
  \begin{tabular}{lcccccccccccc}
    \toprule
     & p@1 & p@5 & p@10 & r@1 & r@5 & r@10 & acc@1 & acc@5 & acc@10 & MAP@1 & MAP@5 & MAP@10 \\
    \midrule
    LRT & 0.371 & 0.105 & 0.054 & 0.016 & 0.038 & 0.052 & 0.017 & 0.033 & 0.059 & 0.017 & 0.029 & 0.042 \\
    Rank-GeoFM & 0.42 & 0.152 & 0.069 & 0.019 & 0.046 & 0.067 & 0.021 & 0.051 & 0.077 & 0.021 & 0.045 & 0.058 \\
    GT-SEER & 0.462 & 0.179 & 0.099 & 0.021 & 0.058 & 0.076 & 0.047 & 0.088 & 0.126 & 0.047 & 0.071 & 0.093  \\
    TA-PLR & 0.457 & 0.184 & 0.096 & 0.025 & 0.062 & 0.087 & 0.045 & 0.101  & 0.143  & 0.045 & 0.074 & 0.091 \\
    STT & 0.547 & 0.209 & 0.125 & 0.032 & 0.071 & 0.09 & 0.055 & 0.137 & 0.174 & 0.055 & 0.084 & 0.098 \\
    \textbf{GEmodel} & 0.598 & 0.224 & \textbf{0.152} & 0.057 & 0.085 & 0.108 & 0.083 & 0.169 & \textbf{0.201} & 0.083 & 0.124 & 0.130 \\
    \textbf{STES} & \textbf{0.606} & \textbf{0.227} & 0.147 & \textbf{0.064} & \textbf{0.089} & \textbf{0.109} & \textbf{0.105} & \textbf{0.176} & 0.199 & \textbf{0.105} & \textbf{0.128} & \textbf{0.132} \\
    \bottomrule
  \end{tabular}
\end{table}

\textbf{STT}: Counts of latent regions and topics are both 150.

\textbf{GT-SEER}: Distance threshold is $1km$; 50 unchecked locations, $\beta/\alpha$ is 0.5; embedding size is 200.

\textbf{TA-PLR}: $\alpha$ is 1; $\lambda$ is 0.01; embedding size is 200.

\textbf{Rank-GeoFM}: $\alpha, \gamma, \epsilon, C, K$ are 0.15, 0.0001, 0.3, 1, 100; 300 nearest neighbors.

\textbf{LRT}: Latent $d$ is 10; $\alpha, \beta, \lambda$ are 2, 2, 1.

\textbf{GEmodel}: Embedding size is 100; 150 samples; the time interval is 30 days.

\textbf{STES}: Embedding size is 200; 15 most preferred venue categories; decay parameters $a_1, a_2, b_1, b_2$ are 0.4, 0.025, 1, 0.95.

We find that our STES algorithm and the GEmodel produce very similar results, which are both significantly better than those of all other baseline methods. GEmodel marginally beats our algorithm in terms of precision and accuracy at top-10 recommendations while in all other cases our STES model is slightly better. However, GEmodel was specifically designed for location recommendation. Although the embeddings can also be utilized in other domains, it is less flexible and general than our algorithm. In the rest of this paper, we will show how our algorithm can be gracefully generalized to other tasks without any adjustments.
% We will show in the furt that our algorithm, without adjustments generalizes gracefully to other tasks and data. Both are properties that GEmodel lacks.

We confirm the statistical significance of performance differences between our method and all of the contesting baselines using McNemar's test~\cite{mcnemar1947note}. The largest mid-p-value is $1.53\times10^{-4}$. 

With respect to \textbf{RQ2}, the results demonstrate the effectiveness of our embedding model and the STES algorithm in user/place characterization for location recommendation. As geographic and temporal aspects are considered in these six approaches in different stages, we argue that our improvement mainly comes from the embedding of venues' functional roles. In addition to indicating \textit{where} and \textit{when} someone is, the functional information further explains \textit{why} someone is there at that time, essentially revealing a person's activity preference beyond specific location preference. Consequently, we attain better relative modelling power when a user is in a region which is away from his/her frequently visited area and literal check-ins at unique locations cannot be levied. Moreover, calculating the \textit{mean} of check-in vectors further leverages the continuity and smoothness of the embedding model and thus establishes more latent correlations between users and locations.    

\subsection{Urban Functional Zone Study}
Previous work has shown that dividing a city into different functional zones is a straightforward yet informative way to define urban areas. The central information according to which to partition functional zones are the inhabitants' interactions with urban spaces. Therefore, we conduct research in this aspect to examine model efficiency in describing people's activities and characterizing places. To do so, we exclusively utilize \textit{feature word} embeddings which contain second-level venue categories and check-in timestamps. We train the embedding model on neighborhood level and represent each neighborhood using the mean of all contained check-in vectors. Then, we implement K-Means clustering on neighborhoods as suggested by \cite{noulas2011exploiting} and \cite{zhu2015learning}.

Remember that in location recommendation, we compared our model with a baseline algorithm GEmodel. Similar to our method, GEmodel generates timestamp and venue category embeddings as well. As an additional comparison, we also characterize neighborhoods with the mean of inner check-in time and venue category vectors trained by GEmodel. 

Zhu \textit{et al.}~\cite{zhu2015learning} demonstrate an effective neighborhood characterization based on the normalized counts of demographic, temporal and spatial aspects of visits. Noulas \textit{et al.}~\cite{noulas2011exploiting} show that the functional zones can be reliably clustered if neighborhoods are represented only by the number of visits at each venue category. Corresponding to these two approaches, we first propose two ground-truth alternatives: (1). l2-normalized counts of \textit{feature words}; (2). l2-normalized counts of venue categories. 

To determine the most qualified ground-truth in our work, we examine the cluster assignments derived from the alternatives using Silhouette Index (SI) \cite{rousseeuw1987silhouettes}, which measures how compact clusters are by computing the average intra-cluster and inter-cluster distances. The SI ranges between -1 for incorrect clustering and +1 for highly dense and well separated clustering. An SI around 0 indicates overlapping clusters.

We increase the count of clusters from 3 to 10 and report the Silhouette indices of clusterings in Table \ref{tab:si}.

\begin{table}
\centering
\caption{Silhouette Index measurements}
\label{tab:si}
\begin{tabular}{lcccccccc}
\hline
 & 3 & 4 & 5 & 6 & 7 & 8 & 9 & 10 \\
\hline
our model & 0.315 & \textbf{0.557} & \textbf{0.415} & 0.233 & 0.208 & 0.252 & 0.244 & 0.354 \\
GEmodel & 0.297 & \textbf{0.453} & \textbf{0.492} & 0.285 & 0.214 & 0.259 & 0.236 & 0.296 \\
alternative 1 & 0.253 & \textbf{0.487} & \textbf{0.443} & 0.271 & 0.209 & 0.237 & 0.211 & 0.323  \\
alternative 2 & 0.281 & \textbf{0.429} & \textbf{0.443} & 0.231 & 0.294 & 0.242 & 0.198 & 0.236 \\
\hline
\end{tabular}
\end{table}          

We can see that all compared methods peak in performance at four or five clusters. Compared with GEmodel based clustering, our embedding model produces more well-defined clusters. Similarly, ground-truth alternative 1 performs better than alternative 2. 

\begin{figure}
\centering
\subfigure[POI-based clustering]{
\label{fig:POI based clustering}
\includegraphics[width=0.35\textwidth]{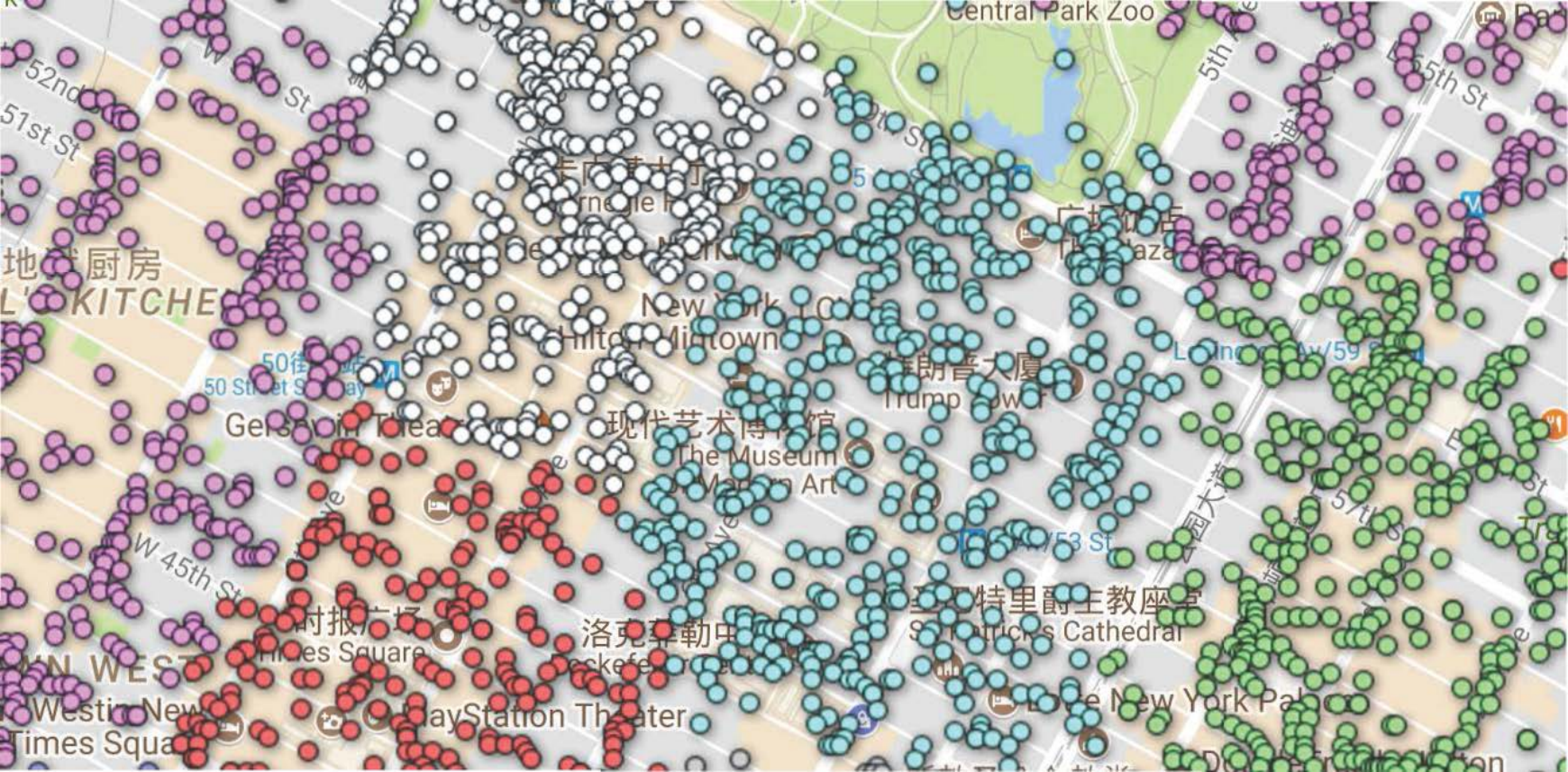}}
\subfigure[Neighborhood-based clustering]{
\label{fig:neighborhood based clustering}
\includegraphics[width=0.35\textwidth]{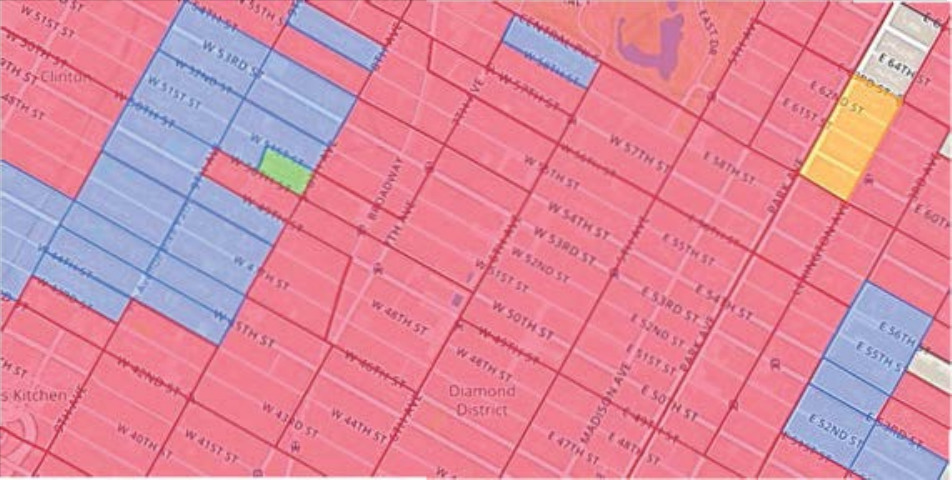}}
\caption{POI-based clustering utilizes check-in information more directly. However, on a higher geographic abstraction level, this results in a globally less representative clustering. On the other hand, since neighborhoods are Census Block Groups defined by the U.S. Census Bureau, neighborhood-based clustering ensures a natural interpretation of the urban functionality and the daily interaction between people and the their surroundings.}
\label{fig:point_or_area}
\end{figure}

We begin by calculating vector representations of neighborhoods and utilize those as the smallest units for clustering. Therefore, our scenario is different from the work described by Cranshaw \textit{et al.}~\cite{cranshaw2012livehoods} in which clustering is based on individual POIs. POI-based clustering leverages check-in information in a more direct manner. However, as demonstrated in Figure~\ref{fig:point_or_area}, it is difficult to utilize POI-based clustering results for  global study of urban functionality. Another two urban clustering works~\cite{yuan2015discovering, zhu2015learning} are excluded from the comparison since both of them require more detailed traces of personal mobility, such as GPS trajectories, which are not available in our dataset. Specifically, they rely on complete travel logs in a period of time during which consecutive leaving and arrival locations and times are recorded. In our scenario, however, check-ins are rather sparse and do not allow for robust computation of such models. 
% We also re-implemented their method. Overall, they can achieve higher Silhouette indices (above 0.8). However, POIs are extremely unevenly distributed into clusters——about 90\% of them are assigned to the first cluster when there are 4-5 clusters.

In the following, we focus on our embedding model based clusters and utilize the alternative 1 (l2-normalized counts of \textit{feature words}) as the ground-truth. 

Given this ground-truth, a more objective validation of the results can be obtained by comparing the clusters derived from the embedding model with those from the ground-truth, aiming for them to be as similar as possible \cite{zhu2015learning}. A common metric for this scenario is the Rand Index (RI) \cite{rand1971objective}. This metric penalizes pair-wise disagreeing cluster assignments across models. In our work, we employ the Adjusted Rand Index (ARI) \cite{milligan1985examination}, which further discounts for expected clustering coherence due to chance.

\begin{figure}
\scriptsize
\centering
\includegraphics[width=0.7\textwidth]{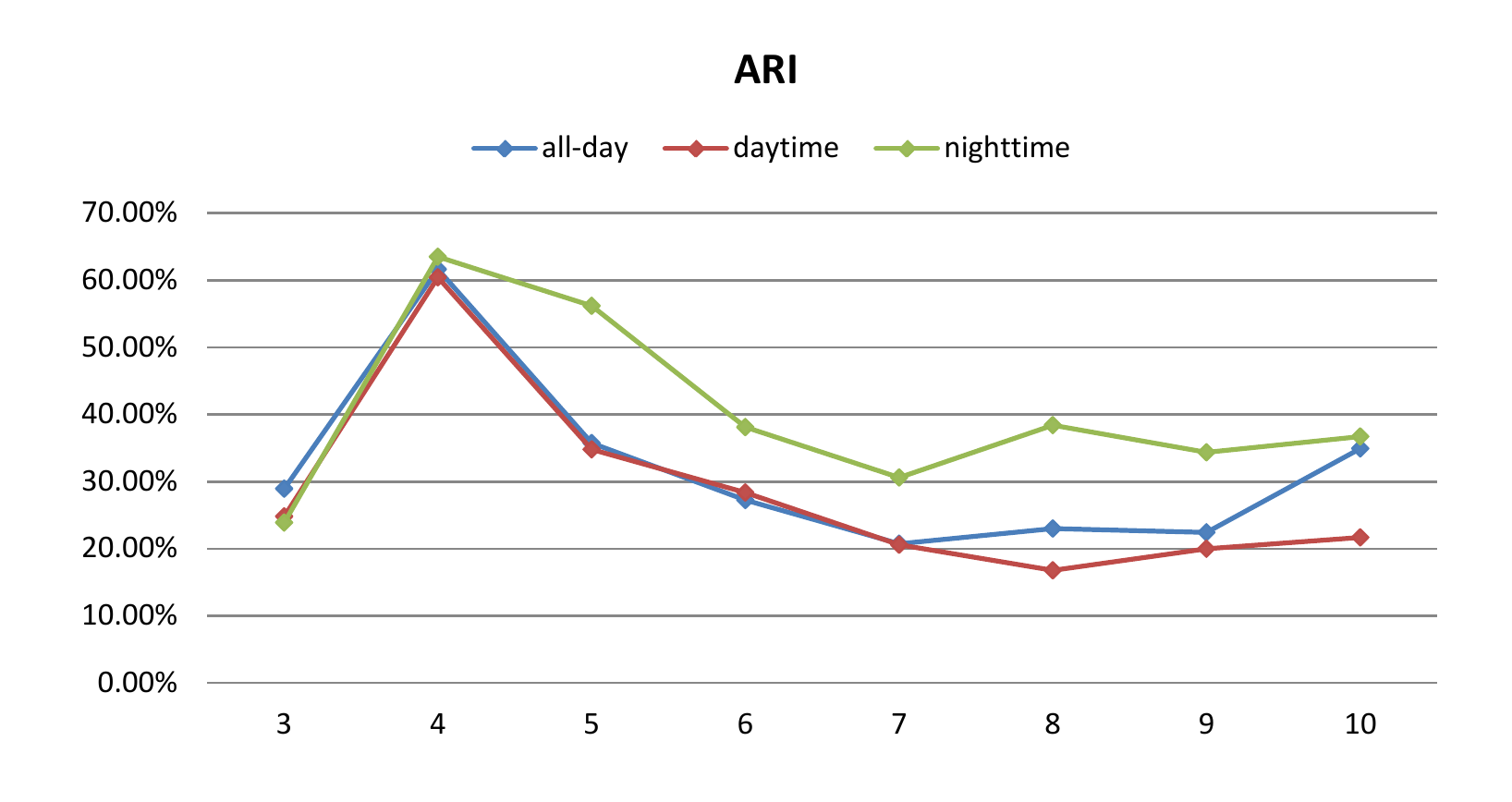}
\caption{Adjusted Rand Index (ARI) of embedding model based clustering and ground-truth based clustering. In all of the cases, ARI peaks at around 61\% with four clusters.}
\label{fig:ARI}
\end{figure}

ARI is bounded in $[-1,1]$, where $1$ corresponds to a perfect match score and random (uniform) assignments lead to a score close to $0$. In addition to the clustering based on all check-ins, we further perform clustering based on only daytime (weekday and weekend morning, noon, and afternoon) check-ins and only nighttime (weekday and weekend evening and night) check-ins. Figure~\ref{fig:ARI} illustrates the ARIs, showing all three cases peaking at four clusters, incidentally the same point as demonstrated by SI in Table~\ref{tab:si}. 
% We compare the clustering memberships of neighborhoods in Figure \ref{fig:clustering}, which visually shows a high degree of overlap.

\begin{figure}
\scriptsize
\centering
\includegraphics[width=0.75\textwidth]{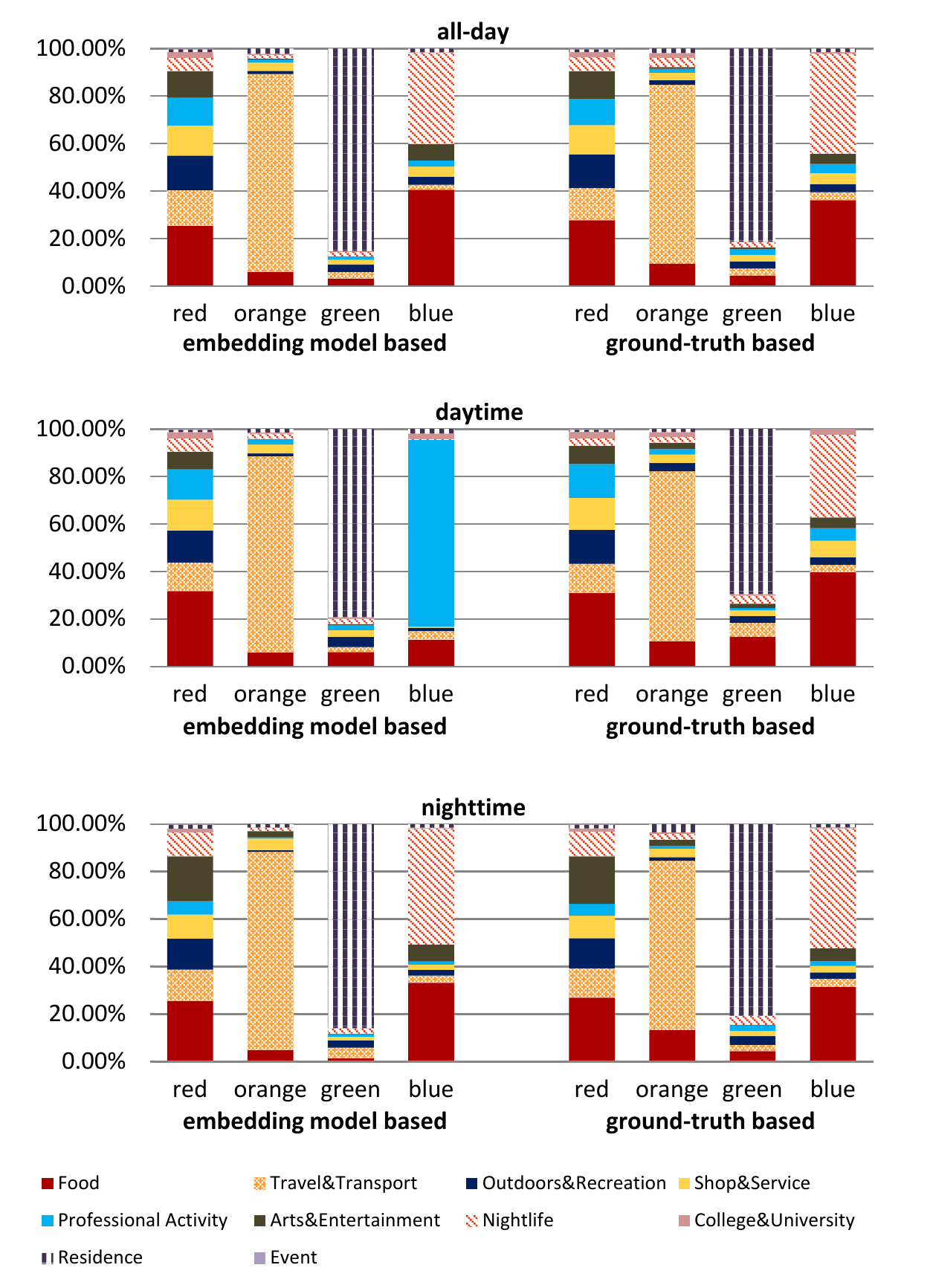}
\caption{Functional composition of clusters. The red cluster represents multi-functional zones; the orange cluster is travel\&transporation dominant and the green cluster is for residence. The blue cluster is mainly for nightlife and food at nighttime; during daytime, it represents professional area for embedding model based clustering while remaining as food-nightlife place for ground-truth based clustering.}
\label{fig:comp}
\end{figure} 

As a consequence, the following results and evaluations are exclusively based on the four-cluster setting. We interpret results semantically as the clustering is based on concrete functionality. In Figure~\ref{fig:comp}, we plot the composition ratio of venue categories in each cluster based on person-time check-ins. As can be expected from the previously observed high ARI, no matter whether all-day, daytime, or nighttime clustering, clusters based on the embedding model and the ground-truth show high similarities in their compositions and have one-to-one correspondence. 

\begin{figure}
\centering
\subfigure[cluster example1]{
\label{fig:cluster example1}
\includegraphics[width=0.65\textwidth]{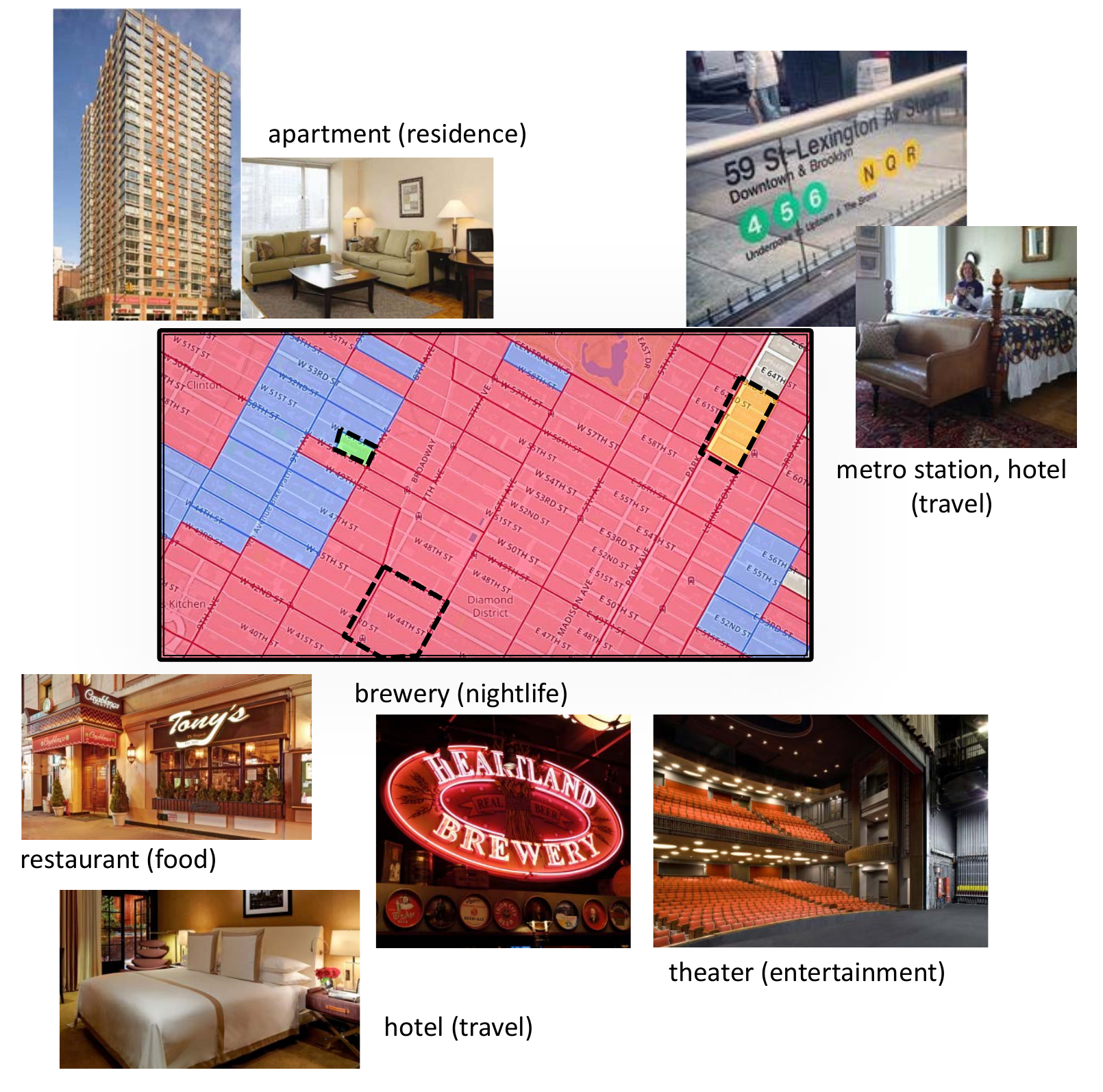}}
\subfigure[cluster example2]{
\label{fig:cluster example2}
\includegraphics[width=0.65\textwidth]{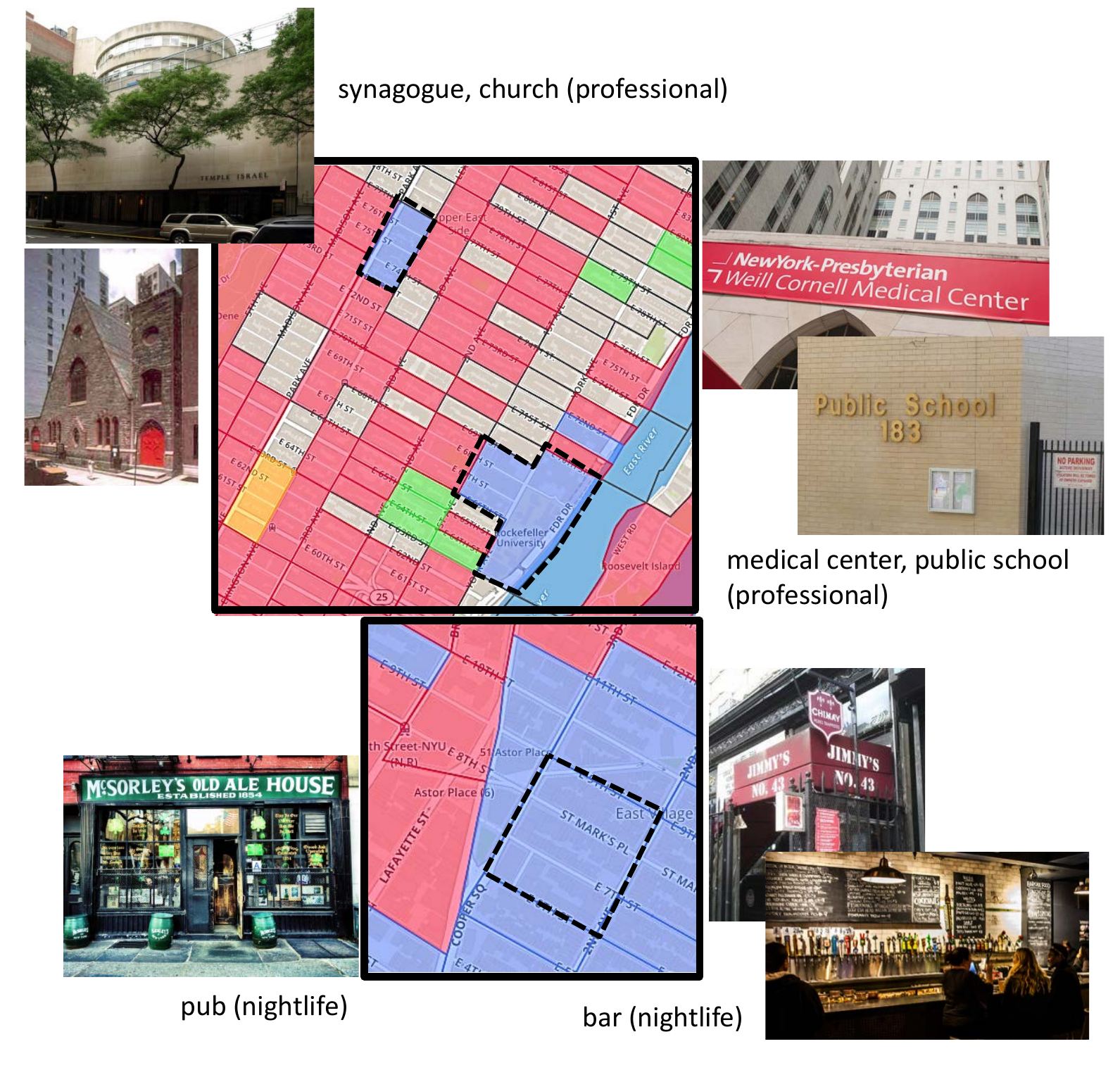}}
\caption{Cluster examples.}
\label{fig:example}
\end{figure}

Let us focus on the embedding model based clustering. We find that in most cases, a cluster has a single dominant functionality. For better visualization, Figure~\ref{fig:example} depicts some cluster examples. In addition, Table~\ref{tab:ratios} lists exact cluster ratios in the five NYC boroughs. In all cases, the red cluster represents multi-functional zones. Except for daylong daily routines, they contain more professional activities during daytime while experiencing more entertainment at nighttime. Geographically, during the day, red neighborhoods are mainly distributed in Manhattan, Brooklyn, and Queens, which are the most densely populated boroughs involving economical, political, and cultural activities. The orange clusters are primarily for travel and transportation. By examining individual neighborhoods, they contain major expressways, main transportation junctions such as Lexington Avenue Station and the Coney Island Complex. The green cluster represents residential zones. Staten Island, The Bronx, and Queens (especially upper Queens) have higher proportions of the green cluster, which is reasonable as these districts display considerable amounts of residential areas. 

\begin{table}
% \scriptsize
\centering
\caption{Cluster ratio in NYC boroughs}
\label{tab:ratios}
\begin{tabular}{clccccc}
\hline
 & & Staten Island & Manhattan & The Bronx & Brooklyn & Queens \\
\hline
\multirow{4}{*}{all-day} & red & 58.3\% & 57.1\% & 56.6\% & 48.8\% & 59.9\% \\ 
    % \cline{2-7}
    & orange & 0\% & 3\% & 23.6\% & 9\% & 9.1\% \\
    % \cline{2-7}
    & green & 30\% & 5.8\% & 14.2\% & 9.9\% & 12.0\% \\
    % \cline{2-7}
    & blue & 11.7\% & 34.1\% & 5.6\% & 32.3\% & 19.0\% \\
\hline
\multirow{4}{*}{daytime} & red & 67.8\% & \textbf{88.7\%} & 58.5\% & \textbf{77.3\%} & \textbf{75.9\%} \\ 
    % \cline{2-7}
    & orange & 0\% & 3.2\% & 22.6\% & 10.2\% & 9.7\% \\
    % \cline{2-7}
    & green & 28.5\% & 6.2\% & 14.2\% & 10.0\% & 11.4\% \\
    % \cline{2-7}
    & blue & 3.7\% & 1.9\% & 4.7\% & 2.5\% & 3\% \\
\hline
\multirow{4}{*}{nighttime} & red & 54.4\% & 54.9\% & 54.3\% & 47.9\% & 55.1\% \\ 
    % \cline{2-7}
    & orange & 0\% & 2.6\% & 24.5\% & 9.1\% & 9.6\% \\
    % \cline{2-7}
    & green & \textbf{33.3\%} & 6.0\% & \textbf{16.0\%} & 10.8\% & \textbf{14.2\%} \\
    % \cline{2-7}
    & blue & 12.3\% & \textbf{36.5\%} & 5.2\% & \textbf{32.2\%} & 21.1\% \\
\hline
\end{tabular}
\end{table}

By comparison, the blue cluster is assigned different functionalities in different timeslots. Generally speaking, its main roles are related to food and nightlife. During daytime, it is predominately used for professional activities, which include work, education, medical service, spiritual activities, and so forth. Some representative places include New York University Langone Medical Center, Ravenswood Generating Station, and Junior High School 217 Robert A Van Wyck. At nighttime, nightlife is the strongest functional component, followed by food. Manhattan and Brooklyn have the largest proportions of blue neighborhoods, and in fact, most of the blue cells are placed in lower Manhattan and northwestern Brooklyn, where in addition to various restaurants, there also exist many popular bars and pubs like McSorley's Old Ale House, the Bridge Cafe, and the Ear Inn.

\begin{figure}
\centering
\subfigure[neighborhood in the day]{
\label{fig:neighborhood in the day}
\includegraphics[width=0.8\textwidth]{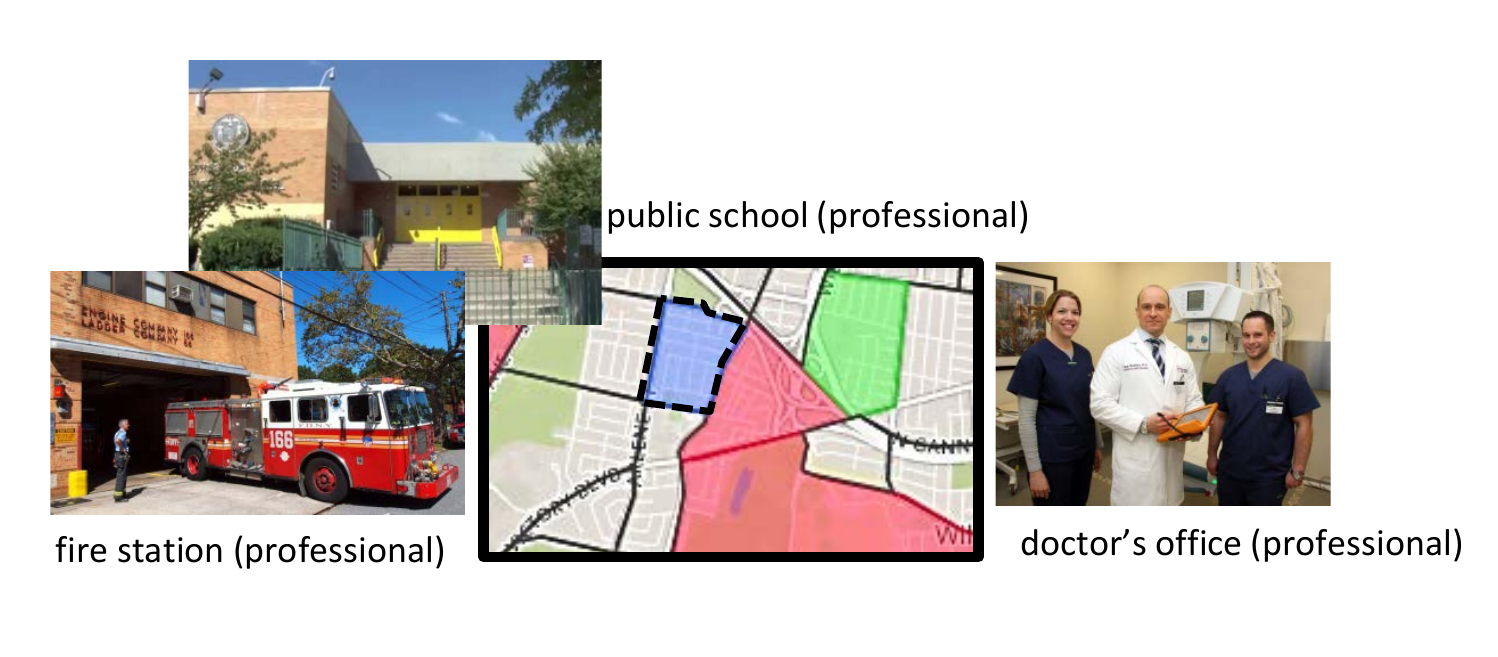}}
\subfigure[neighborhood in the night]{
\label{fig:neighborhood in the night}
\includegraphics[width=0.8\textwidth]{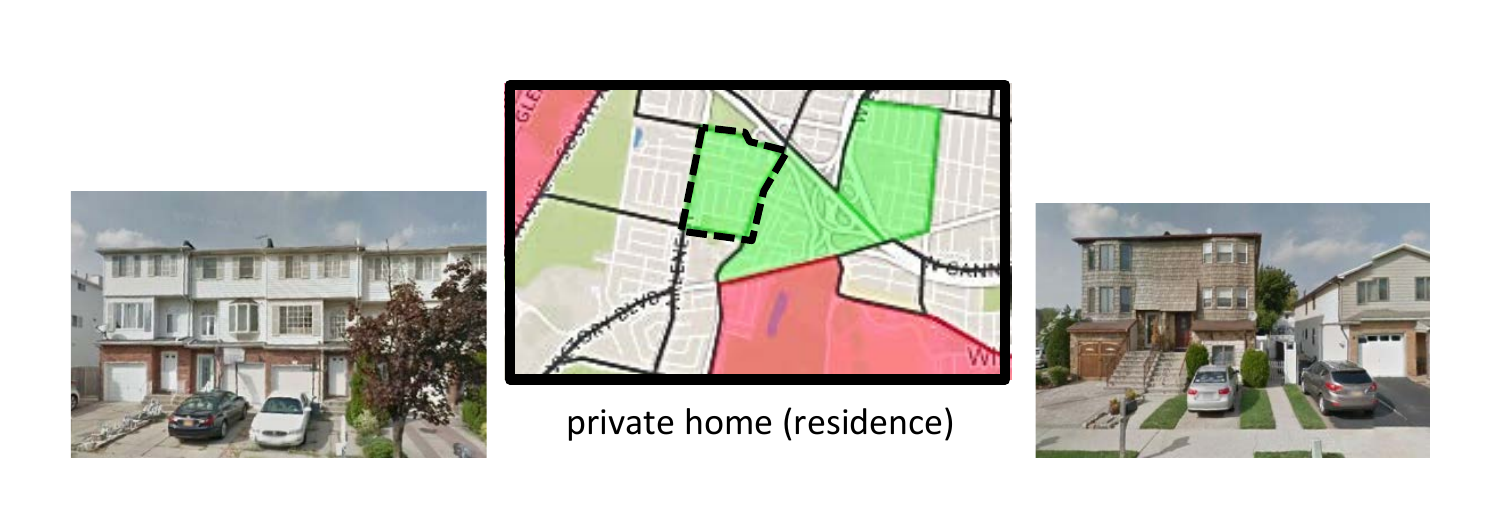}}
\caption{A neighborhood with different functionalities during the day and night. Professional activities are dominant in the day while residence is the main functionality at night.}
\label{fig:neigh}
\end{figure}

This section discusses \textbf{RQ3}. Remember that the embedding model based clustering is purely based on daily check-ins; therefore, it reflects how the city is \textit{actually} used by people. For instance, as the most densely populated area in NYC, Manhattan also boasts some of the city's main transportation hubs and residential buildings. However, orange and green neighborhoods are seldom allocated there. This indicates that the downtown area is generally popular for all types of activities. Moreover, locally popular activities may be time-variant. Figure \ref{fig:neigh} shows a neighborhood in daytime and nighttime. In daytime, it is mainly used for professional activities, dominated by a clinic, a fire station, and a public school. At nighttime, it is mainly residential, which is plausible as most of the buildings in this area are private homes.

\subsection{Crime Prediction}\label{sec:crime}
By providing spatio-temporal embeddings for user and location characterization, our model represents a proxy for the social interactions observed in an urban area. In this section, we will further investigate this capability by addressing a well-known social science problem: crime prediction. Previous work~\cite{van1999time, weisburd2004trajectories} demonstrates that the occurrence of criminal activities is correlated with place types and time, which are both encoded in our check-in embeddings. 

Similar to our urban functional zone study, neighborhoods rather than users are our study subjects in crime prediction. We utilize \textit{feature word} embeddings to characterize neighborhoods. This modification is motivated by the fact that \textit{location words} are only locally descriptive. Therefore, a neighborhood cannot be described without prior training data from the exact location, but such new neighborhood prediction is possible if only modelled with the universally applicable \textit{feature word} vectors, as long as visited venues have the same functional roles. NYC is still the representative city for study and the crime data originates from the NYC Open Data portal\footnote[5]{https://data.cityofnewyork.us/Public-Safety/NYPD-7-Major-Felony-Incidents/hyij-8hr7}.   

% Following that in urban functional zone study, we also represent neighborhoods with timestamp and category embeddings trained by GEmodel and utilize this as

\subsubsection{Crime Rate Prediction}

Let us begin by defining the task of future crime rate prediction~\cite{gerber2014predicting, wang2012automatic} as predicting the next-month crime rate of a neighborhood. We characterize each neighborhood monthly using the mean of check-in vectors in that month, and we assign crime incidents to neighborhoods according to their location coordinates. A neighborhood is labeled as ``Low'' crime rate (occurrences$=$0/month/neighborhood), ``Medium'' crime rate (0$<$occurrences$<$3/month/neighborhood), or ``High'' crime rate (occurrences
$>=$3/month/neighborhood). Averaging over neighborhoods in each month, 31.3\% are of ``Low'', 37.2\% are of ``Medium'', and 31.5\% are of ``High'' and the variance is 0.047 across months. Averaging across months per location, a neighborhood has a 34.1\% chance of ``Low'', a 39.3\% chance of ``Medium'', and a 26.6\% chance of ``High'' with a variance of 0.087.

We take all data points from Mar.\ 2010 to Oct.\ 2010 (9983 neighborhoods) as a training set while those in Nov.\ and Dec.\ 2010 (2151 neighborhoods) represent our test set. 

We define two performance baselines following our urban functional zone study. The first one is to represent neighborhoods with timestamp and category embeddings trained by GEmodel, and the second one is the l2-normalized monthly counts of \textit{feature words}. 
% To define performance baseline, same as that in urban functional zone study, we use l2-normalized monthly counts of \textit{feature words} as the baseline features. 
Remember that we also reviewed several prediction schemes in the literature part, however, they are not directly applicable in our case due to the lack of tweet texts and demographic statistics such as \textit{education levels}.
In addition to these two baselines, we further define a ``random" baseline which refers to the results of consistently predicting the most likely label based on the ground truth of the training data.  

\begin{figure}
% \centering
\subfigure[crime rate prediction accuracy]{
\label{fig:crime rate}
\includegraphics[width=0.45\textwidth]{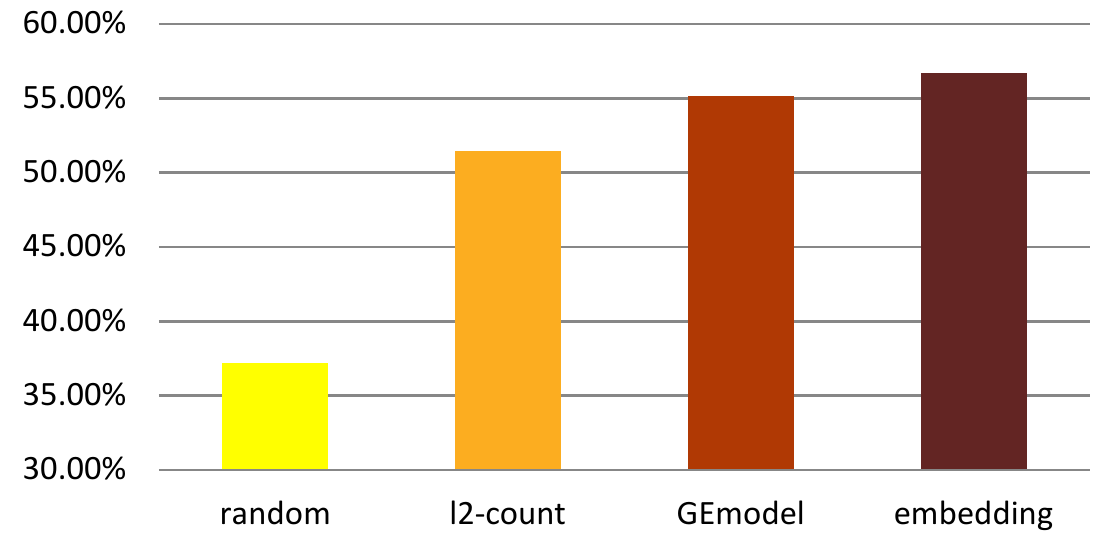}}
\subfigure[crime rate prediction $F_1$ score]{
\label{fig:crime rate f1}
\includegraphics[width=0.45\textwidth]{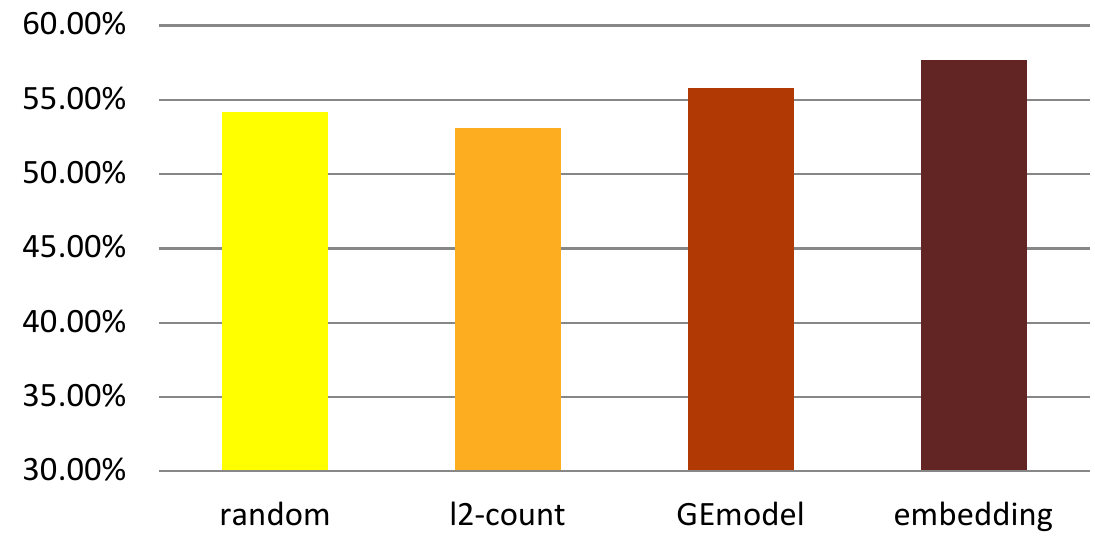}}
\subfigure[crime occurrence prediction accuracy]{
\label{fig:crime occurrence}
\includegraphics[width=0.45\textwidth]{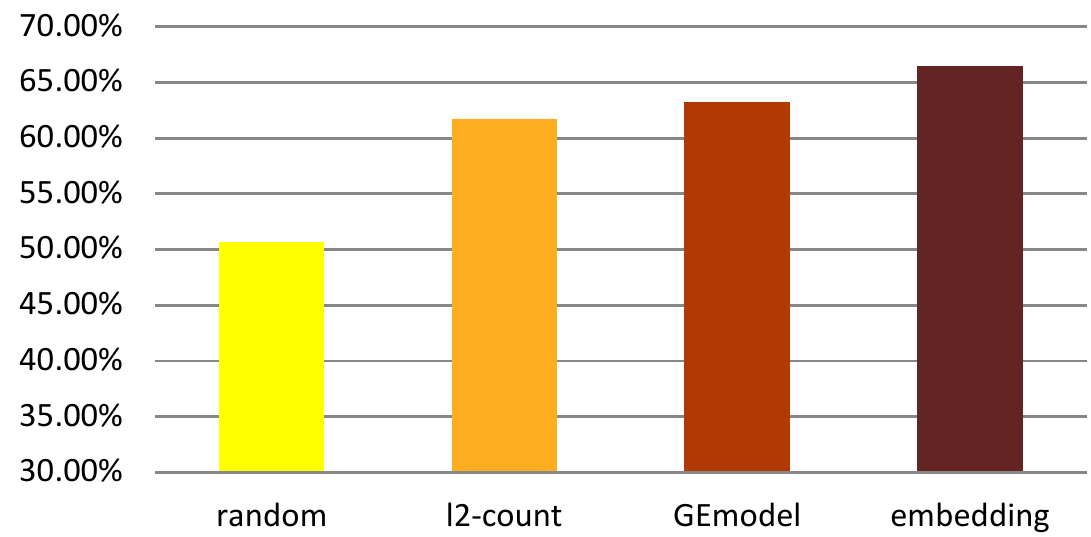}}
\subfigure[crime occurrence prediction $_1$ score]{
\label{fig:crime occurrence f1}
\includegraphics[width=0.45\textwidth]{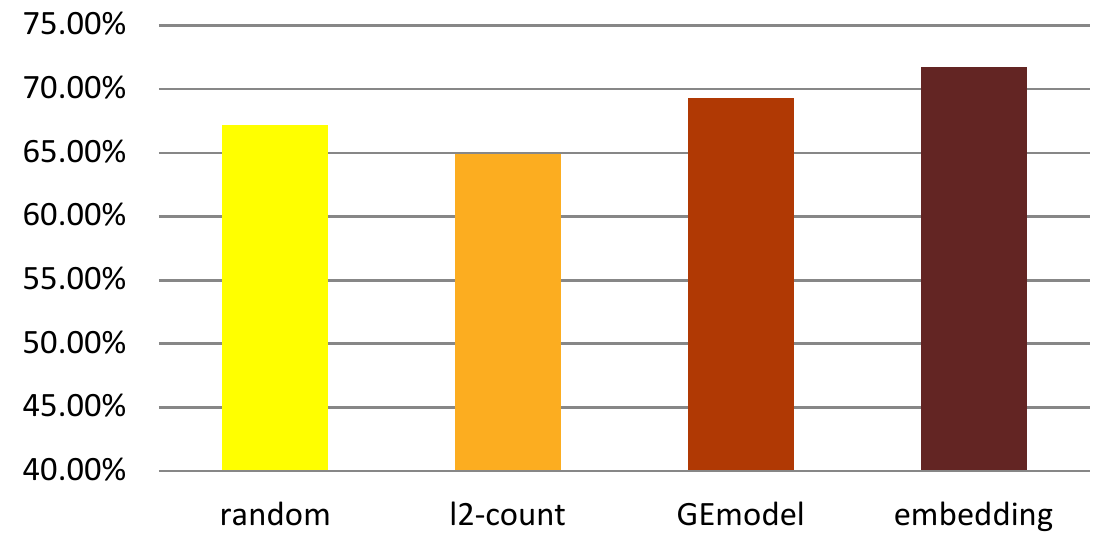}}
\caption{Average crime prediction accuracies and $F_1$ scores in NYC.}
\label{fig:NYC crime}
\end{figure}

We evaluate the performance through accuracy and $F_1$-score and results are shown in Figure~\ref{fig:crime rate} and Figure~\ref{fig:crime rate f1}. After testing various classification frameworks, we use a random forest classifier. In this case the ``random" bar is based on predicting ``Medium" crime rate. 
% The leftmost bar "random" refers to the results from consistently predicting the most likely label based on the ground truth of the training data. 
The figures show that our embedding model produces the best results with significant performance gains over all contestants according to both metrics. We can further note that the $F_1$-score of random baseline is slightly higher than that of the l2-count baseline method, which is mainly due to the high recall in the three-class random guessing case.

% \stepcounter{figure}

\subsubsection{Crime Occurrence Prediction}

Aside from predicting overall crime rates, we are interested in understanding local crime events in greater detail. In our crime data set, a frequently occurring crime type in NYC is \textit{Grand Larceny}. We will now investigate whether we can predict the occurrence of this particular type of crime in a neighborhood within the following month. Experimental settings and classifier choice remain unchanged while the labels are changed into ``No Grand Larceny'' and ``Grand Larceny''. When averaging over neighborhoods in each month, the label ratio is 51\% for ``No Grand Larceny'' and 49\% for ``Grand Larceny'' with a variance of 0.027. For each neighborhood, it has on average a 60.9\% probability that this crime would occur with a variance of 0.082.

Figure~\ref{fig:crime occurrence} and Figure~\ref{fig:crime occurrence f1} demonstrate the average prediction accuracies and $F_1$-scores. Similar to crime rate prediction, we can observe that our model outperforms random guessing as well as both baselines at significance-level. 

Both crime rate and occurrence prediction results pass the McNemar's significance test with the largest mid-p-value of $1.2062\times10^{-7}$. 

\begin{figure}
% \centering
\subfigure[crime rate: time distribution]{
\label{fig:crime rate time}
\includegraphics[width=0.6\textwidth]{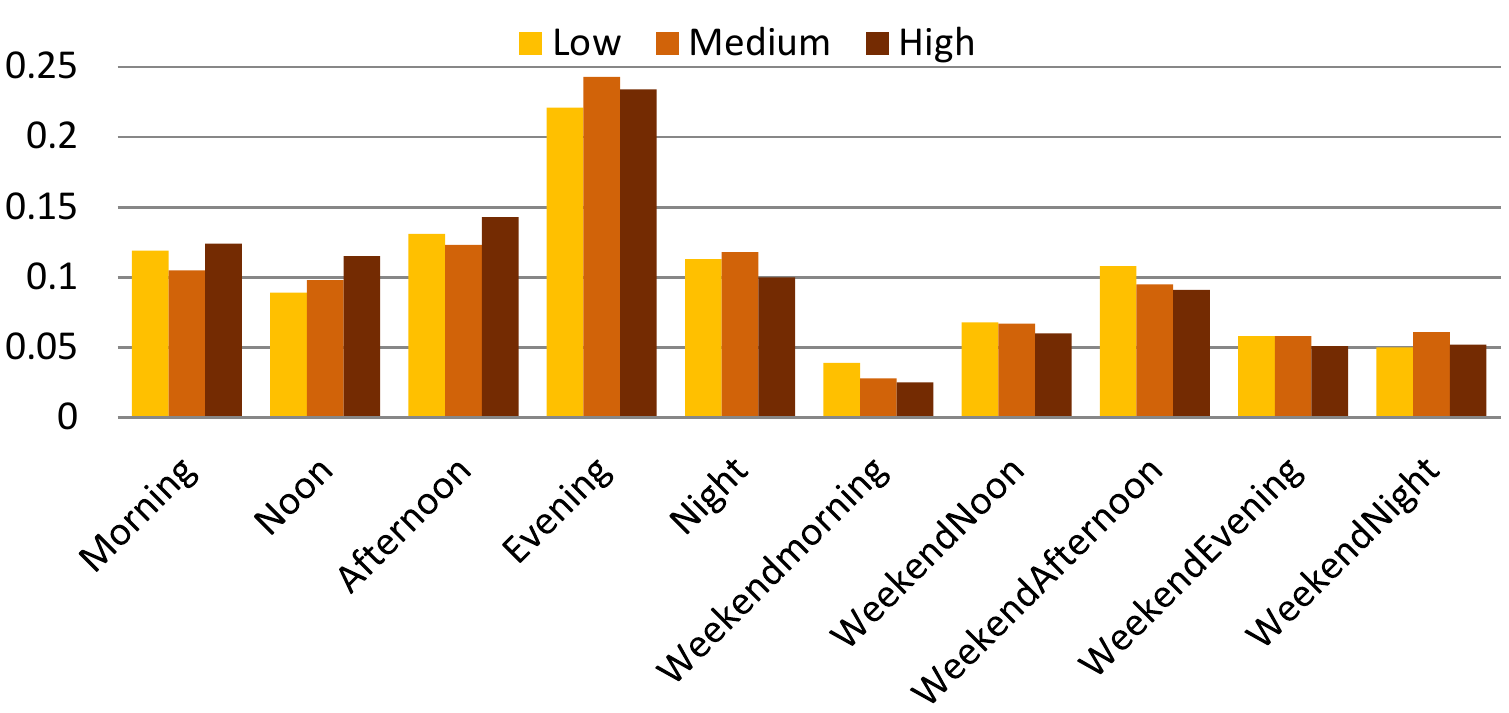}}
\subfigure[crime rate: category distribution]{
\label{fig:crime rate category}
\includegraphics[width=0.6\textwidth]{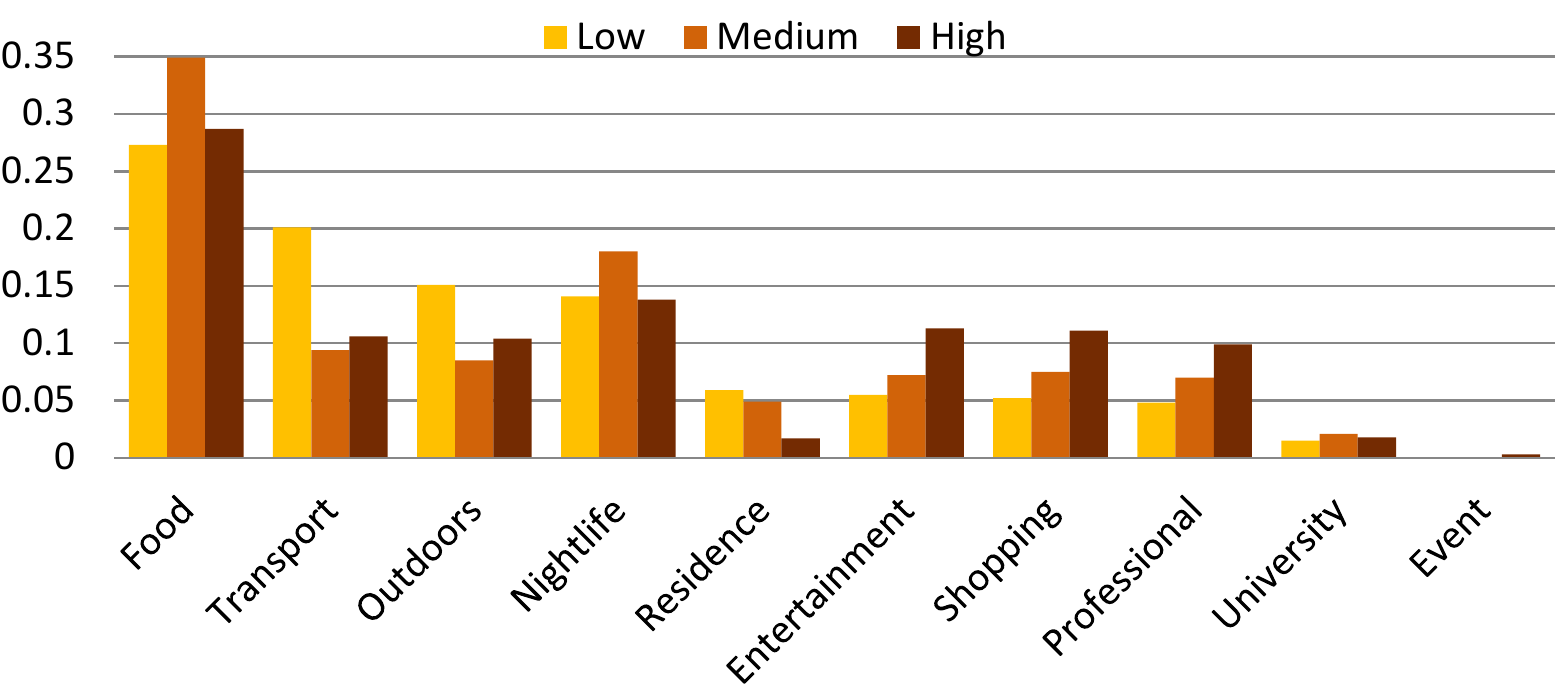}}
\subfigure[crime occurrence: time distribution]{
\label{fig:crime occurrence time}
\includegraphics[width=0.6\textwidth]{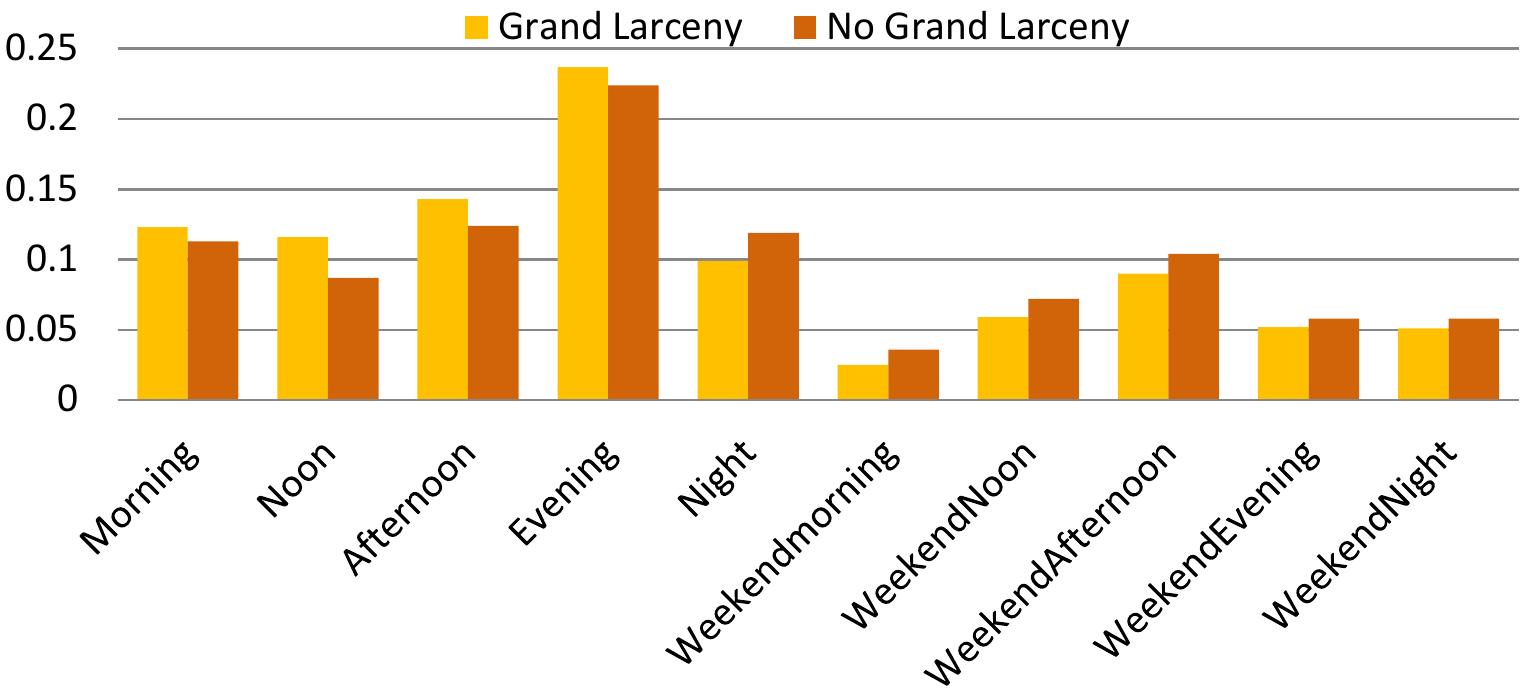}}
\subfigure[crime occurrence: category distribution]{
\label{fig:crime occurrence category}
\includegraphics[width=0.6\textwidth]{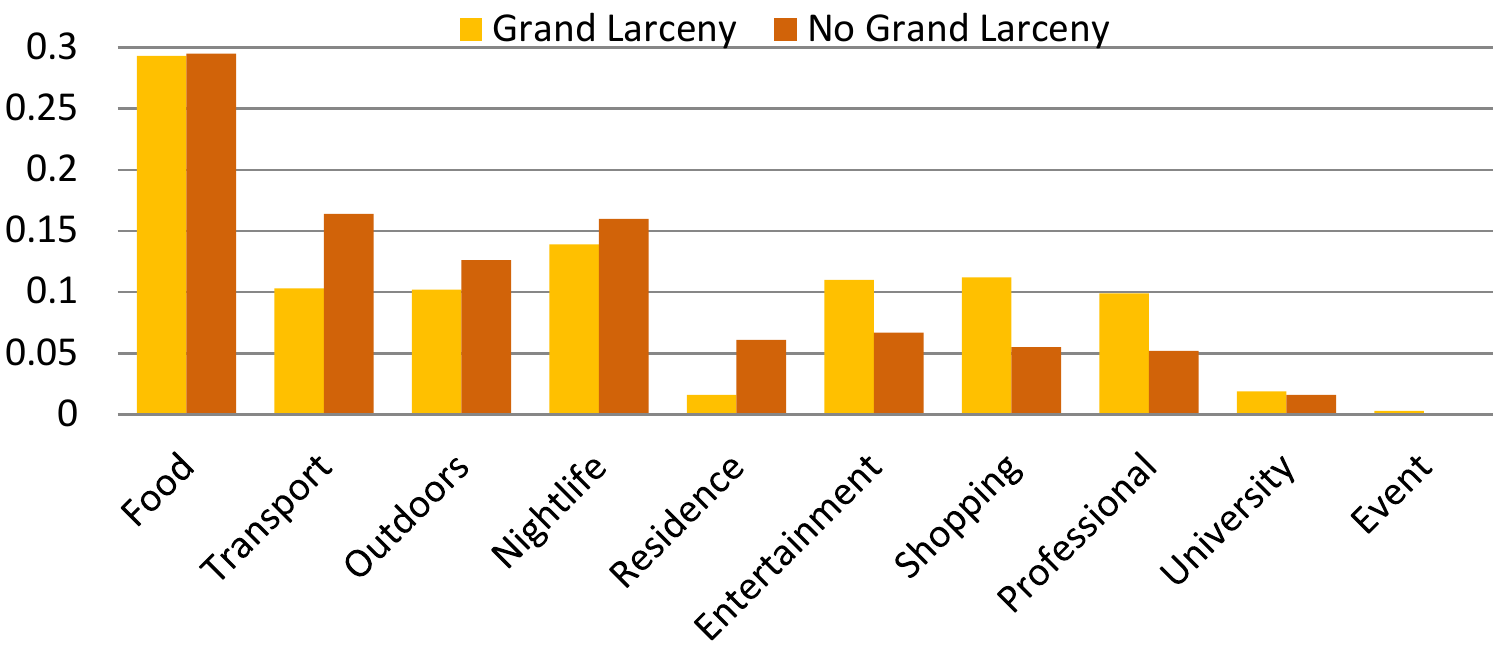}}
\caption{Check-in time and location category distribution in neighborhoods of different labels. }
\label{fig:justification}
\end{figure}

To justify the rationale behind our method and results, we further examine the check-ins in neighborhoods with low/medium/high crime rate and with/without grand larceny, and we plot their check-in time and location category distribution in Figure~\ref{fig:justification}. We can see that neighborhoods with high crime rate are more checked in on weekdays at entertainment (\textit{e.g.} casino), shopping, and professional (\textit{e.g.} business center) places, which also applies to neighborhoods with more grand larceny cases. In response to \textbf{RQ4}, this section demonstrates the latent connection between people's daily activities and crime occurrence in an urban area. The prediction results further demonstrate our embedding model's effectiveness in encoding activity information in place representations, furthering the understanding and inference of social science problems. 

% \stepcounter{figure}

\subsection{Model Generalization}

The training of large-scale embedding models can be a costly process requiring hours or days of computational resources. To save this time, we investigate whether a pre-trained model can be directly applied in other cities while retaining most of its performance. In this way, our approach differs from existing work in~\cite{bao2012location}, that generalizes user profiles across cities (\textit{i.e.}, the user travels) but still requires local modelling in the new city.

In our generalization experiments, NYC is taken as the reference city where the embedding model is trained. With consideration to data size and the geographic distance to NYC, we select Chicago, Los Angeles, Seattle, San Francisco, London, Amsterdam, Bandung, and Jakarta for the generalization test. As before, we delete repeating check-ins from individuals in artificially short time periods and remove both users and locations with less than 10 posts. Eight cities and their check-in statistics are listed in Table~\ref{tab:sta}. We also list NYC statistics for reference. 

% Restricted by the availability of Census Block Group data, we test the model generality on location recommendation and crime prediction only. 

% \stepcounter{table}
\begin{table*}
% \scriptsize
  \caption{Check-In Statistics}
  \label{tab:sta}
  \begin{tabular}{lcccc}
    \toprule
    City  & \# of check-ins & \# of users & \# of locations & avg.\ check-ins/user (density) \\
    \midrule
    Chicago (CH) & 86,117 & 2,755 & 3,678 & 31.26 \\
    Los Angeles (LA) & 118,088 & 4,238 & 5,609 & 27.86 \\
    Seattle (SE) & 44,960 & 1,523 & 2,180 & 29.52  \\
    San Francisco (SF) & 84,494 & 3,285 & 3,605 & 25.72 \\
    London (LO) & 45,270 & 2,182 & 1,922 & 20.75 \\
    Amsterdam (AM) & 49,722 & 1,855 & 1,895 & 26.80 \\
    Bandung (BA) & 23,581 & 1,476 & 996 & 15.98 \\
    Jakarta (JA) & 50,875 & 3,123 & 1,995 & 16.29 \\
    $\ast$ \textit{NYC} & $\ast$ \textit{225,782} & $\ast$ \textit{6,442} & $\ast$ \textit{7,453} & $\ast$ \textit{35.05} \\
    \bottomrule
  \end{tabular}
\end{table*}

\subsubsection{Generalization of Location Recommendation}

Recall that in our earlier investigation in Section~\ref{sub:locRec}, we relied on both \textit{feature words} and \textit{location words} for location recommendation. However, since \textit{location words} are locally descriptive, they cannot be easily taken out of their original frame of reference. As a consequence, we only generalize the NYC-based embedding model for \textit{feature words}, but locally train \textit{location words}.

Upon careful examination, none of the baseline methods can be ported across cities. GT-SEER and TA-PLR methods exclusively focus on local POI embedding; Rank-GeoFM and LRT algorithms rely on location vectorization; STT is dependent on local topic model; GEmodel trains embedding vectors based on bipartite graphs but all of the graphs involve local POIs.  
% as they exclusively focus on the \textbf{local} location id embedding (GT-SEER and TA-PLR), vectorization (Rank-GeoFM and LRT), modelling (STT). 
Therefore, we have to train the baseline methods locally when applying them to different cities.  

As before, we measure performance in terms of \textit{precision}, \textit{recall}, \textit{accuracy}, and \textit{MAP} and the results are listed in Table~\ref{tab:all locs}. Again, GEmodel and STES(local) have very similar performances, and the half-transferred model STES(NYC) produces close and even better results in some cases. For instance, in Seattle, the NYC-based STES model outperforms both GEmodel and the local STES model according to \textit{recall} at top-1 recommendation. This can happen as a consequence of data sparsity since Seattle only offers 45k local check-ins which provide less information than the 226k transferred ones from NYC. 
% Except for Seattle, local STES models consistently perform the best in all the cases in the other seven cities, followed by NYC-based STES and the baseline methods; while in Seattle, NYC-based STES model produces very close results compared with the local STES model and the former outperforms the latter according to \textit{recall} at top-1 recommendation. 

\begin{figure}
% \centering
\subfigure[STES(local) and STES(NYC) gap at A@1]{
\label{fig:diff1}
\includegraphics[width=0.65\textwidth]{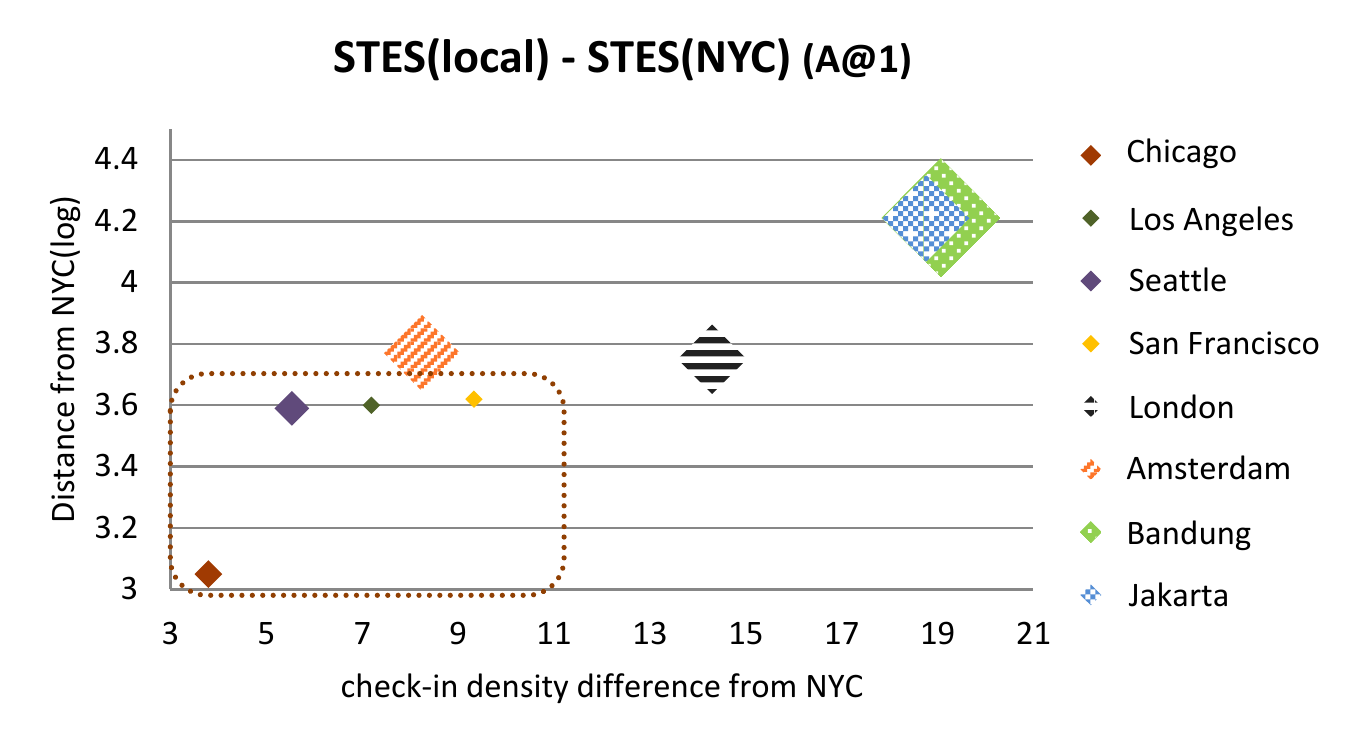}}
\subfigure[STES(local) and STES(NYC) gap at A@5]{
\label{fig:diff2}
\includegraphics[width=0.65\textwidth]{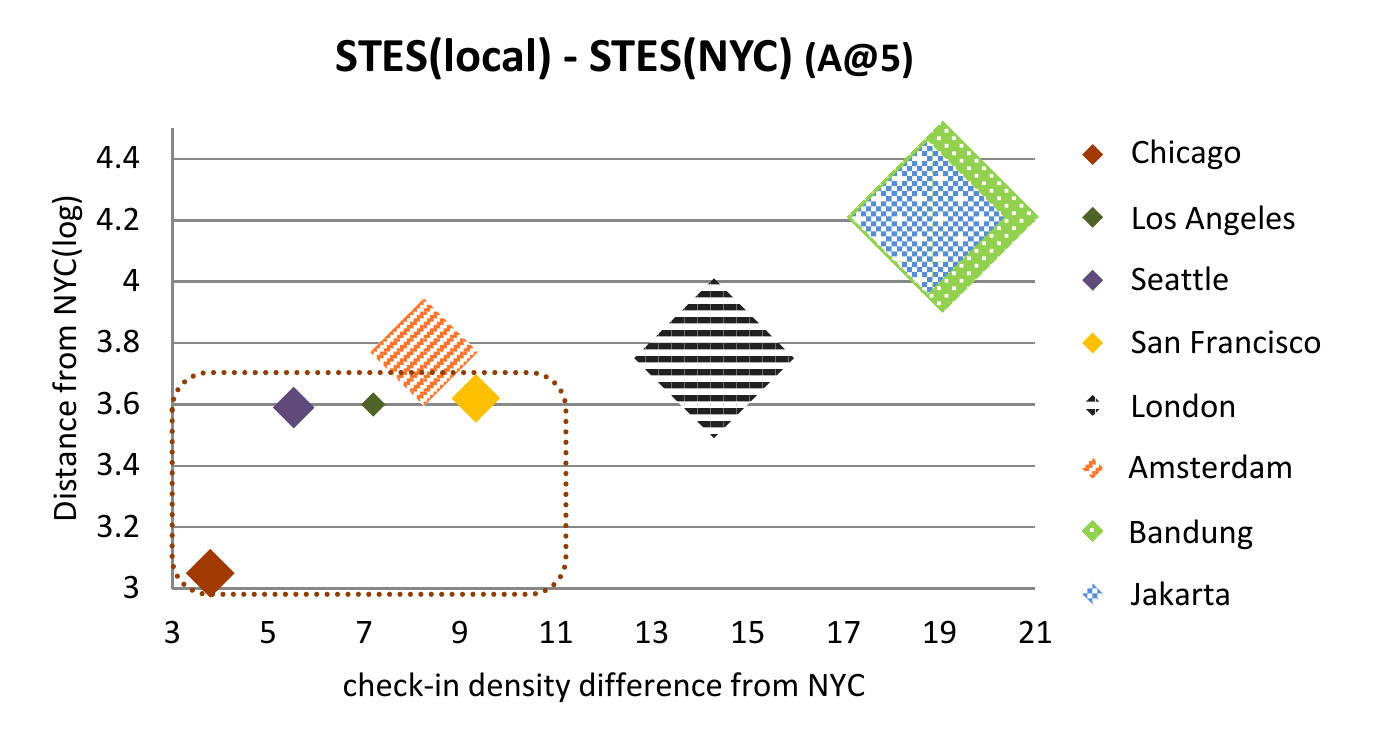}}
\subfigure[STES(local) and STES(NYC) gap at A@10]{
\label{fig:diff3}
\includegraphics[width=0.65\textwidth]{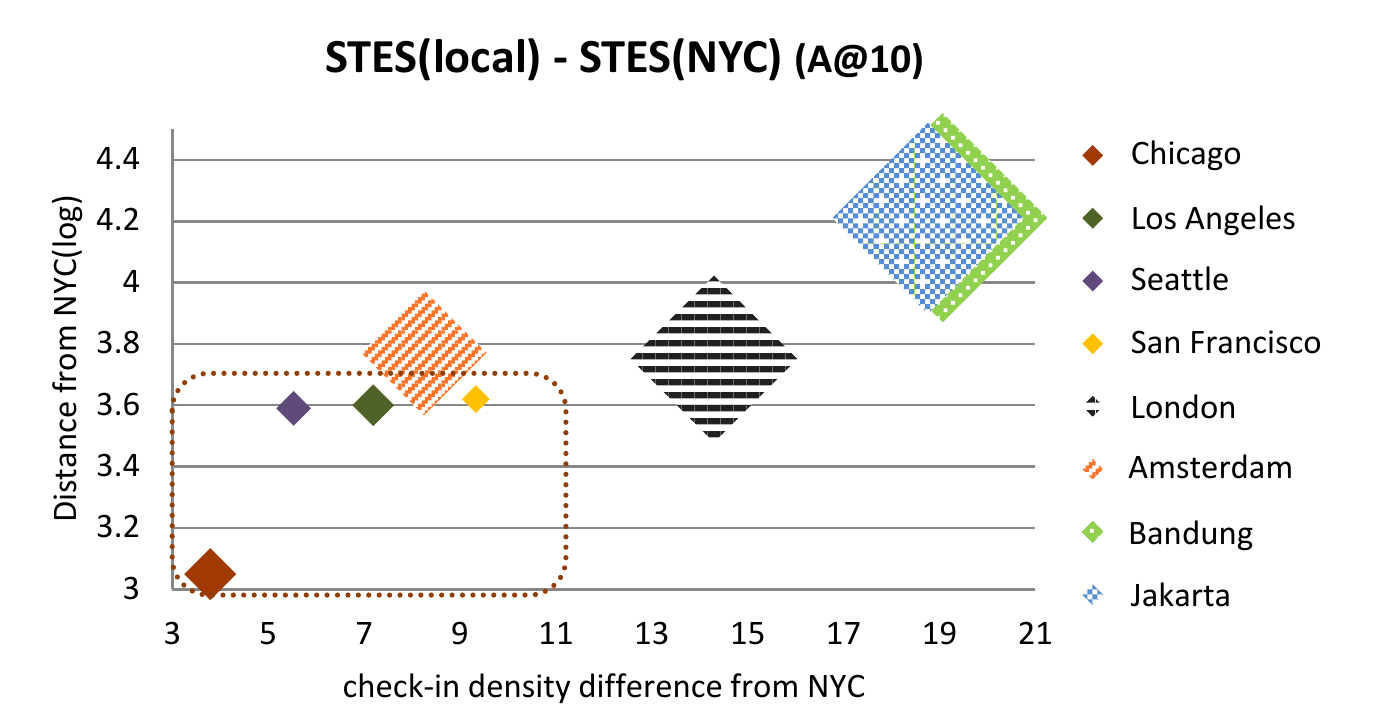}}
\caption{Accuracy differences (represented by rhombus area sizes) between STES(local) and STES(NYC) for top-1, top-5, and top-10 recommendations. X-axis is the check-in density difference from that in NYC and Y-axis is the log distance in km from the city to NYC. U.S.\ cities are framed in a brown dash grid. Accuracy differences become larger with increasing distance and check-in density difference.}
\label{fig:diff}
\end{figure}

Let us now focus on the STES(NYC) and STES(local) models. Upon closer examination, in all of the eight cities, the gaps between STES(NYC) and STES(local) are generally smaller than 3\%. In particular, these two methods almost tie at top-1 recommendation for all U.S.\ cities; while in the other four non-US cities, the NYC-based STES model produces less satisfactory performance with respect to the local models. 
To further understand this observation, we plot the accuracy differences between STES(local) and STES(NYC) in top-1, top-5, and top-10 cases in Figure~\ref{fig:diff}. From the figure, we can see that both the difference in local check-in density and the distance from NYC appear to exert influence on generalization performance.  While the former is less significant, the latter plays a key role in model portability as indicated by comparison among Los Angeles, Amsterdam, and San Francisco. 
We argue that the geographic distance from NYC is a proxy for cultural differences in the way that urban zones are used. Specifically, life style in Southeast Asia is distinct from that in the U.S., which also applies for Europe where the difference seems smaller. Therefore, the NYC-based embedding model is well adapted to other U.S.\ cities but somewhat less competitive in European and Asian cities. 

\subsubsection{Generalization of Crime Prediction}

Due to the lack of comparable publicly available crime statistics, generalization experiments are only conducted for U.S.\ cities. The problem scenario and experimental settings remain unchanged from the description in Section~\ref{sec:crime}. Locally collected and processed\footnote[6]{https://data.cityofchicago.org/Public-Safety/Crimes-2001-to-present/ijzp-q8t2}\footnote[7]{https://data.seattle.gov/Public-Safety/Crimes-2010/q3s4-jm2b}\footnote[8]{http://shq.lasdnews.net/CrimeStats/CAASS/desc.html}\footnote[9]{https://data.sfgov.org/Public-Safety/SFPD-Incidents-from-1-January-2003/tmnf-yvry}, crime types vary among cities. 
Like in NYC, we implement three-grade crime rate prediction with different crime rate thresholds in each city, except for Los Angeles, for which we conduct a two-category experiment since only a very limited amount of crimes were recorded. This is also the reason for the lack of a single frequent crime type in Los Angeles while the other crime data sets report locally frequent crime types.
For brevity's sake, we will focus on \textit{Criminal Damage} for Chicago, \textit{Vandalism} for San Francisco and \textit{Property Damage} for Seattle as they are locally common and classify the neighborhoods more evenly than other crime types. Table~\ref{tab:crime_sta} lists the training set size, test set size, crime rate ratio, and crime occurrence ratio in each city. 

% \stepcounter{table}
\begin{table}
% \scriptsize
  \caption{Crime Statistics in U.S. Cities}
  \label{tab:crime_sta}
  \begin{tabular}{lcccc}
    \toprule
    City & Training & Test & Low:(Medium):High & No:Yes \\
    \midrule
    CH & 4,668 & 964 & 36.7\%:32.1\%:31.2\% & 48.9\%:51.1\% \\
    LA & 5,488 & 1,206 & 59.8\%:40.2\% & NA  \\
    SE & 1,588 & 324 & 33.2\%:33.5\%:33.3\% & 50.4\%:49.6\% \\
    SF & 2,280 & 476 & 31.7\%:38.1\%:30.2\% & 40.0\%:60.0\% \\
  \bottomrule
\end{tabular}
\end{table}

\begin{figure}
% \centering
\subfigure[Crime rate prediction accuracy]{
\label{fig:crime rate all}
\includegraphics[width=0.45\textwidth]{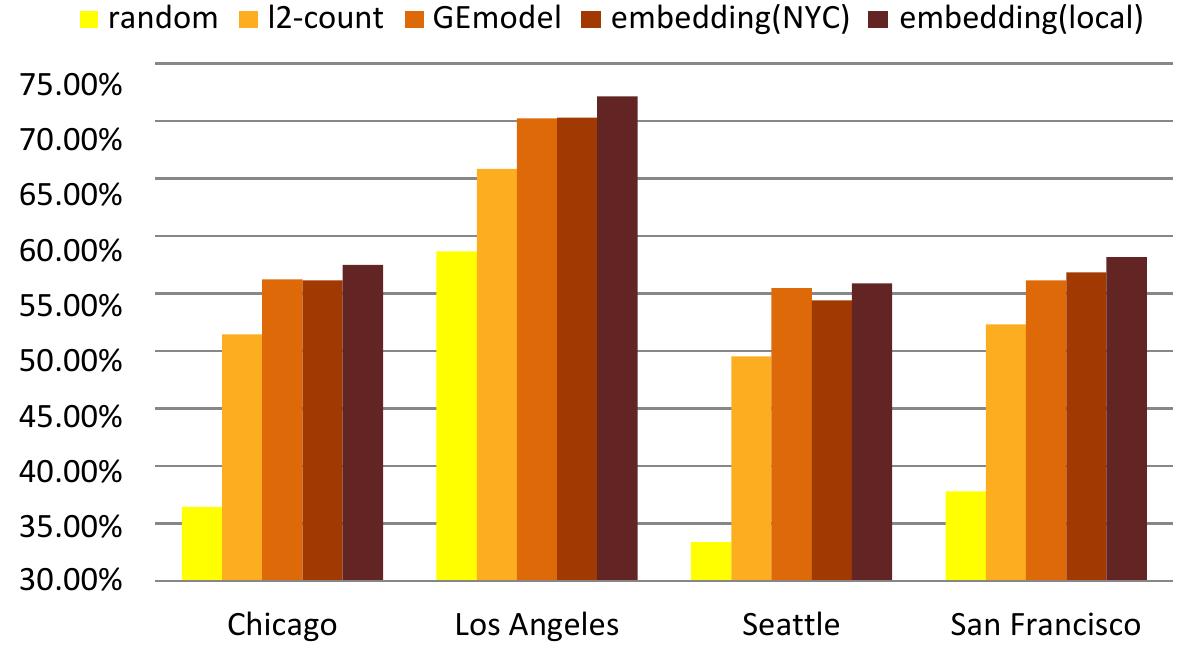}}
\subfigure[Crime rate prediction f1 score]{
\label{fig:crime rate all f1}
\includegraphics[width=0.45\textwidth]{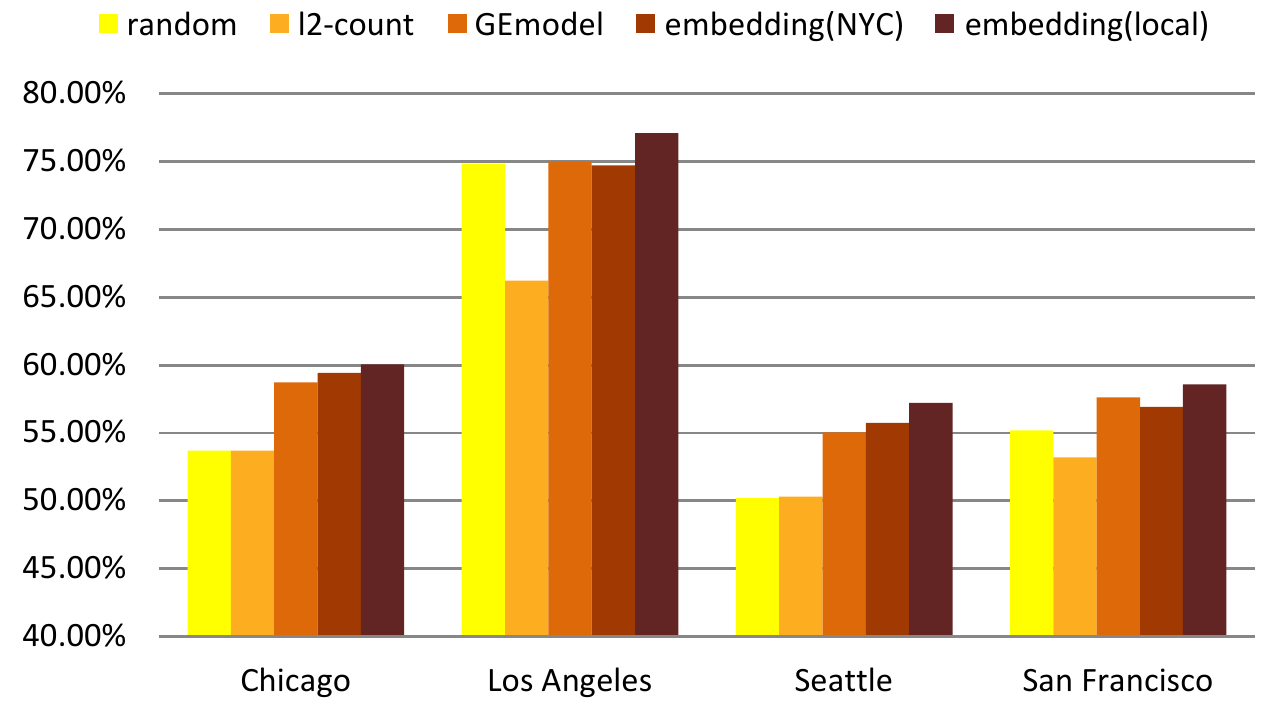}}
\subfigure[Crime occurrence prediction accuracy]{
\label{fig:crime occurrence all}
\includegraphics[width=0.45\textwidth]{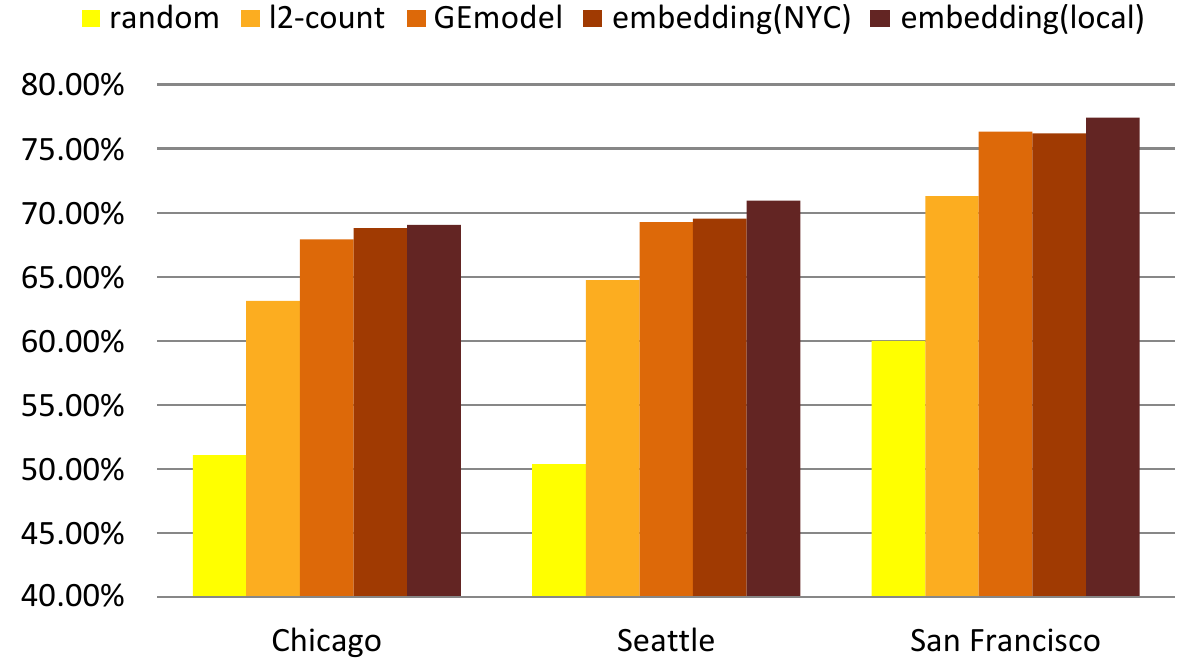}}
\subfigure[Crime occurrence prediction f1 score]{
\label{fig:crime occurrence all f1}
\includegraphics[width=0.45\textwidth]{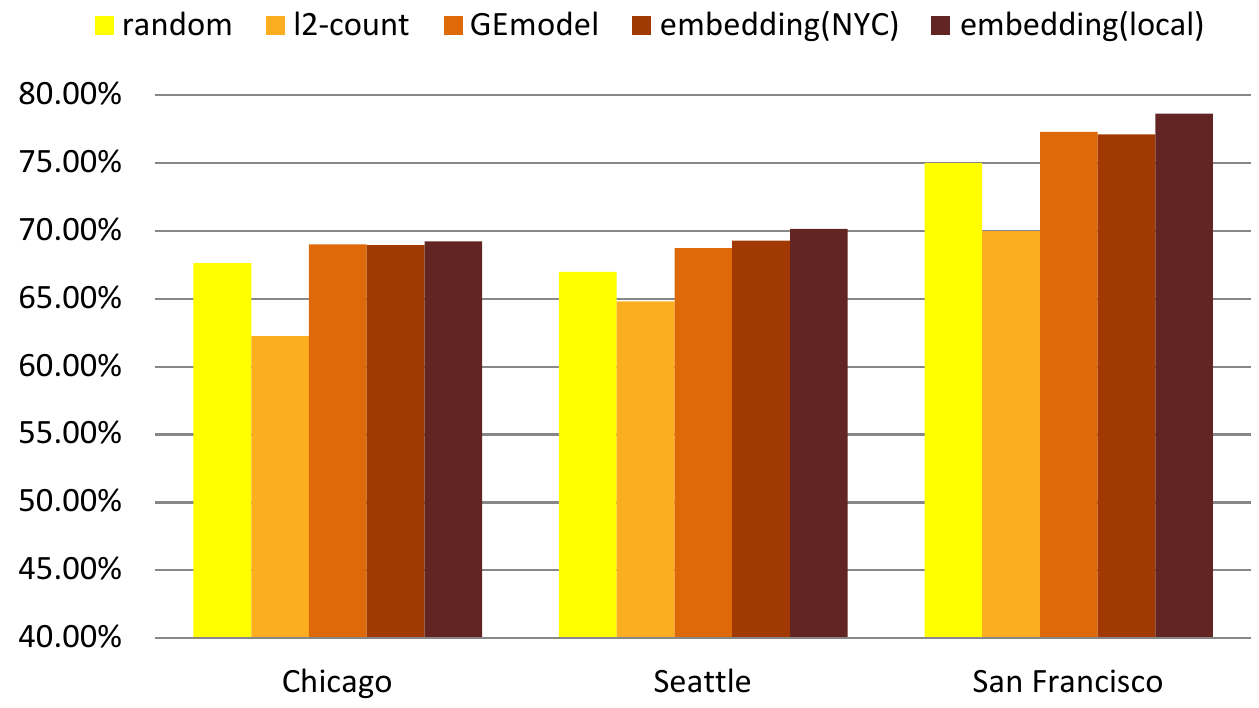}}
\caption{Generalized cross-city crime rate and occurrence prediction in U.S. cities.}
\label{fig:all_crime}
\end{figure}

Figure~\ref{fig:all_crime} shows the prediction results. In terms of accuracy, local embedding models perform best in all cases. NYC-based STES and GEmodels result in comparable performances, falling behind the local STES model by only less than 2\%.
% followed by the NYC-based embedding models and baseline models in sequence.
% In particular, the NYC-based models only fall behind the local model by less than 2\% for crime rate prediction and the gap is even smaller in crime occurrence prediction. 
In terms of $F_1$-scores, local embedding models still outperform all the other methods. NYC-based embedding models and GEmodels lead to similar patterns as those in accuracy evaluation. Random guessing in some cases results in a comparable performance to NYC-based embedding models due to its high recall.

% \stepcounter{figure}

These generalization experiments answer \textbf{RQ5} by demonstrating only mild transfer errors. 
Like before, we conduct McNemar's test and ascertain that the performance improvement from baselines to the embedding model are significant at the level of 0.05. 
This property can remove the need for time-consuming local training while maintaining competitive performance. Additionally, based on the performance differences, we can qualitatively compare similarities between pilot cities and the reference city with respect to their inhabitants' daily activities, life style, and even culture.

% \stepcounter{table}
\begin{table*}
\tiny
  \caption{Location Recommendation Performance in Eight Cities.}
  \label{tab:all locs}
  \begin{tabular}{llcccccccccccc}
    \toprule
     & & p@1 & p@5 & p@10 & r@1 & r@5 & r@10 & acc@1 & acc@5 & acc@10 & MAP@1 & MAP@5 & MAP@10 \\
    \hline
    & LRT & 0.427 & 0.075 & 0.053 & 0.014 & 0.032 & 0.049 & 0.011 & 0.032 & 0.044 & 0.011 & 0.026 & 0.032  \\
    & Rank-GeoFM & 0.501 & 0.118 & 0.066 & 0.023 & 0.047 & 0.061 & 0.019 & 0.052 & 0.065 & 0.019 & 0.038 & 0.054 \\
    & GT-SEER & 0.553 & 0.121 & 0.076 & 0.026 & 0.053 & 0.069 & 0.028 & 0.071 & 0.099 & 0.028 & 0.062 & 0.077 \\
 CH & TA-PLR & 0.672 & 0.196 & 0.128 & 0.037 & 0.065 & 0.082 & 0.079 & 0.144 & 0.182 & 0.079 & 0.108 & 0.117 \\
    & STT & 0.685 & 0.203 & 0.132 & 0.041 & 0.069 & 0.085 & 0.071 & 0.157 & 0.205 & 0.071 & 0.112 & 0.12 \\
    & \textbf{GEmodel} & 0.734 & \textbf{0.262} & \textbf{0.163} & 0.055 & 0.091 & \textbf{0.117} & 0.122 & \textbf{0.206} & \textbf{0.241} & 0.122 & 0.148 & \textbf{0.166}\\
    & STES(NYC) & 0.729 & 0.238 & 0.143 & 0.056 & 0.088 & 0.104 & 0.121 & 0.197 & 0.231 & 0.121 & 0.145 & 0.152 \\
    & \textbf{STES(local)} & \textbf{0.736} & 0.257 & 0.16 & \textbf{0.061} & \textbf{0.094} & 0.112 & \textbf{0.125} & 0.205 & 0.239 & \textbf{0.125} & \textbf{0.151} & 0.163 \\
    \hline
    & LRT & 0.427 & 0.076 & 0.032 & 0.011 & 0.028 & 0.03& 0.017 & 0.049 & 0.062 & 0.017 & 0.035 & 0.048  \\
    & Rank-GeoFM & 0.464 & 0.09 & 0.051 & 0.014 & 0.033 & 0.036 & 0.023 & 0.076 & 0.107 & 0.023 & 0.049 & 0.057 \\
    & GT-SEER & 0.498 & 0.115 & 0.076 & 0.021 & 0.039 & 0.042 & 0.036 & 0.095 & 0.131 & 0.036 & 0.062 & 0.076 \\
 LA & TA-PLR & 0.524 & 0.137 & 0.096 & 0.023 & 0.045 & 0.051 & 0.05 & 0.121 & 0.149 & 0.05 & 0.067 & 0.084 \\
    & STT & 0.523 & 0.149 & 0.102 & 0.024 & 0.048 & 0.056 & 0.053 & 0.123 & 0.154 & 0.053 & 0.071 & 0.082 \\
    & \textbf{GEmodel} & 0.594 & 0.227 & \textbf{0.155} & 0.053 & 0.089 & 0.105 & 0.111 & 0.186 & \textbf{0.227} & 0.111 & 0.135 & 0.138 \\
    & STES(NYC) & 0.61 & 0.231 & 0.15 & 0.055 & 0.086 & 0.099 & 0.112 & 0.185 & 0.215 & 0.112 & 0.131 & 0.137 \\
    & \textbf{STES(local)} & \textbf{0.617} & \textbf{0.232} & 0.154 & \textbf{0.059} & \textbf{0.096} & \textbf{0.111} & \textbf{0.114} & \textbf{0.189} & 0.221 & \textbf{0.114} & \textbf{0.139} & \textbf{0.142} \\ 
    \hline
    & LRT & 0.258 & 0.063 & 0.049 & 0.011 & 0.018 & 0.027 & 0.019 & 0.042 & 0.075 & 0.019 & 0.36 & 0.052 \\
    & Rank-GeoFM & 0.307 & 0.072 & 0.058 & 0.016 & 0.024 & 0.041 & 0.021 & 0.069 & 0.093 & 0.021 & 0.047 & 0.063 \\
    & GT-SEER & 0.334 & 0.085 & 0.063 & 0.019 & 0.038 & 0.046 & 0.039 & 0.082 & 0.113 & 0.039 & 0.059 & 0.076 \\
 SF & TA-PLR & 0.387 & 0.129 & 0.088 & 0.025 & 0.046 & 0.058 & 0.054 & 0.112 & 0.159 & 0.054 & 0.079 & 0.093 \\
    & STT & 0.396 & 0.141 & 0.095 & 0.032 & 0.051 & 0.06 & 0.062 & 0.121 & 0.158 & 0.062 & 0.084 & 0.091  \\
    & \textbf{GEmodel} & \textbf{0.434} & 0.173 & 0.121 & 0.052 & \textbf{0.076} & 0.084 & 0.097 & 0.153 & \textbf{0.187} & 0.097 & \textbf{0.118} & \textbf{0.125} \\
    & STES(NYC) & 0.425 & 0.172 & 0.122 & 0.048 & 0.070 & 0.082 & 0.098 & 0.154 & 0.182 & 0.098 & 0.112 & 0.117 \\
    & \textbf{STES(local)} & 0.432 & \textbf{0.176} & \textbf{0.125} & \textbf{0.054} & 0.075 & \textbf{0.089} & \textbf{0.103} & \textbf{0.162} & 0.186 & \textbf{0.103} & 0.117 & 0.124 \\  
    \hline
    & LRT & 0.452 & 0.088 & 0.057 & 0.013 & 0.019 & 0.048 & 0.01 & 0.063 & 0.079 & 0.01 & 0.024 & 0.046 \\
    & Rank-GeoFM & 0.478 & 0.111 & 0.065 & 0.018 & 0.026 & 0.057 & 0.028 & 0.081 & 0.109 & 0.028 & 0.039 & 0.062 \\
    & GT-SEER & 0.512 & 0.143 & 0.079 & 0.028 & 0.037 & 0.066 & 0.037 & 0.101 & 0.145 & 0.037 & 0.056 & 0.074 \\
SE & TA-PLR & 0.566 & 0.189 & 0.105 & 0.032 & 0.043 & 0.075 & 0.05 & 0.131 & 0.189 & 0.05 & 0.074 & 0.091 \\
    & STT & 0.574 & 0.205 & 0.112 & 0.037 & 0.05 & 0.081 & 0.069 & 0.153 & 0.196 & 0.069 & 0.086 & 0.097 \\
    & \textbf{GEmodel} & 0.639 & 0.247 & \textbf{0.170} & 0.052 & 0.089 & 0.105 & 0.118 & 0.207 & 0.246 & 0.118 & 0.141 & 0.153 \\
    & \textbf{STES(NYC)} & 0.643 & 0.252 & 0.167 & \textbf{0.058} & \textbf{0.091} & 0.104 & 0.121 & 0.205 & 0.242 & 0.121 & 0.143 & 0.15 \\
    & \textbf{STES(local)} & \textbf{0.645} & \textbf{0.257} & 0.169 & 0.054 & \textbf{0.091} & \textbf{0.106} & \textbf{0.124} & \textbf{0.215} & \textbf{0.252} & \textbf{0.124} & \textbf{0.147} & \textbf{0.156} \\ 
    \hline
    & LRT & 0.324 & 0.052 & 0.035 & 0.017 & 0.029 & 0.034 & 0.018 & 0.049 & 0.083 & 0.018 & 0.026 & 0.037  \\
    & Rank-GeoFM & 0.385 & 0.07 & 0.053 & 0.024 & 0.036 & 0.041 & 0.027 & 0.076 & 0.112 & 0.027 & 0.048 & 0.062 \\
    & GT-SEER & 0.397 & 0.089 & 0.054 & 0.027 & 0.042 & 0.049 & 0.041 & 0.094 & 0.131 & 0.041 & 0.065 & 0.078 \\
LO & TA-PLR & 0.428 & 0.124 & 0.069 & 0.038 & 0.056 & 0.061 & 0.073 & 0.158 & 0.182 & 0.073 & 0.101 & 0.112 \\
    & STT & 0.432 & 0.127 & 0.075 & 0.042 & 0.055 & 0.064 & 0.079 & 0.158 & 0.189 & 0.079 & 0.098 & 0.109 \\
    & \textbf{GEmodel} & 0.507 & 0.162 & 0.096 & 0.059 & 0.077 & 0.085 & 0.134 & 0.198 & \textbf{0.246} & 0.134 & 0.156 & 0.162 \\
    & STES(NYC) & 0.501 & 0.159 & 0.097 & 0.051 & 0.07 & 0.079 & 0.131 & 0.193 & 0.221 & 0.131 & 0.147 & 0.153  \\
    & \textbf{STES(local)} & \textbf{0.514} & \textbf{0.169} & \textbf{0.103} & \textbf{0.062} & \textbf{0.081} & \textbf{0.094} & \textbf{0.141} & \textbf{0.216} & 0.245 & \textbf{0.141} & \textbf{0.158} & \textbf{0.163}\\
    \hline
    & LRT & 0.523 & 0.074 & 0.053 & 0.016 & 0.032 & 0.043 & 0.029 & 0.101 & 0.13 & 0.029 & 0.047 & 0.062 \\
    & Rank-GeoFM & 0.605 & 0.119 & 0.068 & 0.023 & 0.047 & 0.056 & 0.042 & 0.123 & 0.153 & 0.042 & 0.074 & 0.08 \\
    & GT-SEER & 0.632 & 0.136 & 0.085 & 0.036 & 0.063 & 0.079 & 0.061 & 0.143 & 0.171 & 0.061 & 0.097 & 0.105 \\
AM & TA-PLR & 0.715 & 0.208 & 0.122 & 0.048 & 0.093 & 0.116 & 0.082 & 0.217 & 0.273 & 0.082 & 0.146 & 0.159 \\
    & STT & 0.702 & 0.195 & 0.116 & 0.048 & 0.089 & 0.123 & 0.101 & 0.205 & 0.252 & 0.101 & 0.154 & 0.163 \\
    & \textbf{GEmodel} & 0.771 & 0.243 & 0.132 & 0.075 & 0.111 & \textbf{0.143} & 0.175 & 0.265 & 0.302 & 0.175 & 0.199 & 0.217 \\
    & STES(NYC) & 0.766 & 0.234 & 0.134 & 0.077 & 0.112 & 0.128 & 0.178 & 0.267 & 0.304 & 0.178 & 0.2 & 0.207  \\
    & \textbf{STES(local)} & \textbf{0.776} & \textbf{0.253} & \textbf{0.139} & \textbf{0.087} & \textbf{0.126} & 0.141 & \textbf{0.189} & \textbf{0.283} & \textbf{0.322} & \textbf{0.189} & \textbf{0.213} & \textbf{0.225} \\ 
    \hline
    & LRT & 0.289 & 0.065 & 0.048 & 0.012 & 0.021 & 0.027 & 0.01 & 0.033 & 0.052 & 0.01 & 0.016 & 0.029  \\
    & Rank-GeoFM & 0.325 & 0.067 & 0.052 & 0.017 & 0.026 & 0.033 & 0.013 & 0.040 & 0.069 & 0.013 & 0.026 & 0.038 \\
    & GT-SEER & 0.421 & 0.086 & 0.059 & 0.022 & 0.034 & 0.041 & 0.023 & 0.062 & 0.087 & 0.023 & 0.035 & 0.058 \\
 JA & TA-PLR & 0.488 & 0.099 & 0.053 & 0.027 & 0.049 & 0.05 & 0.032 & 0.082 & 0.154 & 0.032 & 0.047 & 0.065 \\
    & STT & 0.532 & 0.116 & 0.063 & 0.028 & 0.057 & 0.068 & 0.048 & 0.113 & 0.179 & 0.048 & 0.065 & 0.084 \\
    & \textbf{GEmodel} & 0.586 & 0.168 & \textbf{0.109} & 0.057 & 0.079 & 0.094 & 0.128 & 0.176 & \textbf{0.219} & 0.128 & 0.143 & \textbf{0.167} \\ 
    & STES(NYC) & 0.581 & 0.162 & 0.091 & 0.059 & 0.082 & 0.093 & 0.124 & 0.173 & 0.19 & 0.124 & 0.137 & 0.141  \\
    & \textbf{STES(local)} & \textbf{0.596} & \textbf{0.176} & 0.108 & \textbf{0.062} & \textbf{0.097} & \textbf{0.106} & \textbf{0.135} & \textbf{0.189} & 0.218 & \textbf{0.135} & \textbf{0.151} & 0.165 \\ 
    \hline
    & LRT & 0.405 & 0.053 & 0.037 & 0.015 & 0.028 & 0.034 & 0.012 & 0.028 & 0.046 & 0.012 & 0.02 & 0.028  \\
    & Rank-GeoFM & 0.452 & 0.059 & 0.044 & 0.018 & 0.036 & 0.047 & 0.018 & 0.043 & 0.061 & 0.018 & 0.026 & 0.037 \\
    & GT-SEER & 0.493 & 0.082 & 0.05 & 0.021 & 0.046 & 0.054 & 0.024 & 0.062 & 0.083 & 0.024 & 0.031 & 0.056 \\
 BA & TA-PLR & 0.555 & 0.12 & 0.078 & 0.036 & 0.053 & 0.062 & 0.044 & 0.118 & 0.168 & 0.044 & 0.06 & 0.071 \\
    & STT & 0.583 & 0.165 & 0.102 & 0.052 & 0.067 & 0.089 & 0.061 & 0.134 & 0.182 & 0.061 & 0.082 & 0.097 \\
    & \textbf{GEmodel} & 0.649 & 0.224 & 0.153 & 0.067 & 0.102 & \textbf{0.129} & 0.127 & 0.186 & 0.217 & 0.127 & \textbf{0.164} & \textbf{0.172} \\
    & STES(NYC) & 0.641 & 0.215 & 0.144 & 0.069 & 0.092 & 0.107 & 0.121 & 0.179 & 0.193 & 0.121 & 0.148 & 0.153 \\
    & \textbf{STES(local)} & \textbf{0.662} & \textbf{0.234} & \textbf{0.167} & \textbf{0.079} & \textbf{0.108} & 0.127 & \textbf{0.138} & \textbf{0.207} & \textbf{0.223} & \textbf{0.138} & 0.161 & 0.17 \\ 
    \bottomrule
  \end{tabular}
\end{table*}

\section{Conclusions}\label{sec:conclusion}

In this study, we propose an unsupervised embedding model that learns to represent social network check-ins based on their functional, temporal, and geographic aspects in the form of dense numerical vectors in a semantic space. Item correlations are well captured in terms of activity and location similarities.  

We show three model applications: \textit{location recommendation}, \textit{urban functional zone study}, and \textit{crime prediction}. Our embedding model based recommendation algorithm STES outperforms a wide range of state-of-the-art methods according to four performance metrics. Urban functional zone study shows us intuitive patterns about people's activities in different urban areas. Crime prediction demonstrates the model's effectiveness at capturing location properties and verifies the possibility to infer crime rates or occurrences from residents' daily activities. Furthermore, we confirm that the embedding model has good generality and can be trained in well-represented cities before being applied in other places with only a small generalization error.

In particular, we compare our model with a competitive algorithm GEmodel. Both our method and GEmodel are based on neural network techniques considering temporal, geographic, and functional aspects. In the three application domains, GEmodel produces very similar results to our algorithm, especially in location recommendation. However, GEmodel was designed to be trained with more information such as visitors' reviews which are not available in some cases. In addition, GEmodel needs to be trained locally as their embeddings of regions, timestamps, and words are updated together with the local POI embeddings. Our proposed model does not require any information beyond what is available from common location-based service APIs.

There are several interesting lines of future investigation: (1). While the current model has been designed to be easily trained with minimal restrictions regarding task and data, we are interested in incorporating it in task-specific end-to-end architectures to attain further performance improvements.
% We may also include some other messages during the embedding training process, \textit{e.g.}, users' demographic information and social relations, to describe places and users from more perspectives.
(2). We are interested to explore whether incorporating explicit geographic coordinates can generate more descriptive location embeddings than merely relying on venue IDs. Such embeddings can, for instance, be trained using location neighborhoods which are determined by their coordinates.
(3). On the application side, since the model was designed as a general-purpose descriptor, it has the potential to be employed in a broad range of social tasks such as household income prediction or travel time inference.

% Appendix
% \appendix
% \section{Switching times}
% \section{Supplementary materials}

\begin{acks}
We thank Zhiyuan Cheng for providing us with the check-in data set and Bo Hu for sharing their location recommendation algorithm details. We also thank Jan D\"orrie, Zack Zhu, Hangxin Lu, and Hongyu Xiao for their insightful discussions. 

\end{acks}

% Bibliography
% \bibliographystyle{ACM-Reference-Format}
% \bibliography{sample-bibliography}


\end{document}